%% file: StructureFunctions_v7.5.2.tex
\documentclass[aps,prd,twocolumn,superscriptaddress]{revtex4-1}
\pdfoutput=1
\usepackage{graphicx}
\usepackage{amssymb}
\usepackage{amsmath}
\usepackage{epstopdf}
\usepackage{xcolor}
\usepackage{lineno}
\usepackage{tabularx}
\usepackage{multirow}

\newcommand{\xs}{$x$ }

\newcommand{\xq}{$(x,Q^2)$}

\begin{document}

\title{Nuclear Structure Functions at a Future Electron-Ion Collider}

\author{E.C.~Aschenauer}\email{elke@bnl.gov}
\affiliation{Physics Department, Brookhaven National Laboratory, \break Upton, NY 11973, U.S.A.}
\author{S.~Fazio}\email{sfazio@bnl.gov}
\affiliation{Physics Department, Brookhaven National Laboratory, \break Upton, NY 11973, U.S.A.}
\author{M.~A.~C.~Lamont}\email{maclamont@gmail.com}
\affiliation{Physics Department, Brookhaven National Laboratory, \break Upton, NY 11973, U.S.A.}
\author{H.~Paukkunen}\email{hannu.paukkunen@jyu.fi}
\affiliation{University of Jyvaskyla, Department of Physics, P.O. Box 35, 
             FI-40014 University of Jyvaskyla, Finland. 
             }
\affiliation{Helsinki Institute of Physics, P.O. Box 64, FI-00014 University of Helsinki, Finland} 
\author{P.~Zurita}\email{mzurita@bnl.gov}
\affiliation{Physics Department, Brookhaven National Laboratory, \break Upton, NY 11973, U.S.A.}

\begin{abstract}
The quantitative knowledge of heavy nuclei's partonic structure is currently limited to rather 
large values of momentum fraction $x$ -- robust experimental constraints below $x \sim 10^{-2}$ at 
low resolution scale $Q^2$ are particularly scarce. This is in sharp contrast to the free proton's 
structure which has been probed in deep inelastic scattering (DIS) measurements down to 
$x \sim 10^{-5}$ at perturbative resolution scales. The construction of an Electron-Ion Collider 
(EIC) with a possibility to operate with a wide variety of nuclei, will allow one to explore 
the low-$x$ region in much greater detail. In the present paper we simulate the extraction of the 
nuclear structure functions from measurements of inclusive and charm reduced cross sections at an 
EIC. The potential constraints are studied by analyzing simulated data directly in a next-to-leading 
order global fit of nuclear parton distribution functions based on the recent EPPS16 analysis. A 
special emphasis is placed on studying the impact an EIC would have on extracting the nuclear gluon 
PDF, the partonic component most prone to non-linear effects at low $Q^2$. 
In comparison to the current knowledge, we find that the gluon PDF can be measured at an EIC with significantly reduced uncertainties.

\end{abstract}

\date{\today}

\maketitle


\input{MasterTexFiles/Introduction}

\input{MasterTexFiles/EIC}

\input{MasterTexFiles/Observables}

\input{MasterTexFiles/MonteCarlo}
\input{MasterTexFiles/nPDFfit}

\input{MasterTexFiles/Conclusions}

\begin{acknowledgments}
We are very grateful to the EIC group at BNL whose ongoing efforts made this analysis possible.
E.C.A., S.F., and P.Z. acknowledge the support by the U.S. Department of Energy under
contract number No. DE-SC0012704.
The work of H.P. is currently supported by the Academy of Finland, Project 308301. In addition, H.P. acknowledges the funding from the Academy of Finland, Project 297058; the European Research Council grant HotLHC ERC-2011-StG-279579 ; Ministerio de Ciencia e Innovaci\'on of Spain and FEDER, project FPA2014-58293-C2-1-P; Xunta de Galicia (Conselleria de Educacion) --- he has been part of the Strategic Unit AGRUP2015/11.

\end{acknowledgments}

\input{MasterTexFiles/References}
 \end{document}

%% file: MasterTexFiles/Introduction.tex
\section{Introduction}
\label{sec:Introduction}

The deep-inelastic-scattering (DIS) experiments at the HERA collider have yielded versatile, very accurate information on the partonic structure of the free proton in a wide kinematic range~\cite{Abramowicz:2015mha} and contributed significantly to the theoretical advances in the sector of Quantum Chromodynamics (QCD).
The reach in Bjorken's \xs -- the fraction of longitudinal momentum of the nucleon carried by the parton -- goes almost down to $10^{-5}$ in the region of high four-momentum transfer $Q^2 \gtrsim 1 \, {\rm GeV}^2$, where the perturbative QCD (pQCD) is applicable.
The HERA experiments performed several measurements of neutral and charged-current reactions~\cite{Abramowicz:2015mha}, as well as jet \cite{Andreev:2016tgi} and heavy-flavor cross sections \cite{Abramowicz:1900rp}. These data, in varying combinations, form
the backbone of all the modern global fits of free-proton parton distribution functions (PDFs) \cite{Forte:2013wc,Rojo:2015acz}. In turn, reliable PDFs are a crucial ingredient in interpreting the measurements in hadron colliders like the Relativistic Heavy-Ion Collider (RHIC) at the Brookhaven National Laboratory (BNL) and the Large Hadron Collider (LHC) at the European Organization for Nuclear Research (CERN).  
High-precision PDFs are also indispensable to distinguish signals from processes beyond the Standard Model.

Notwithstanding the remarkable phenomenological success of QCD, 
a detailed understanding of the partonic structure of bound nuclei is still lacking.
In the collinear factorized approach to pQCD, these particles are described by nuclear parton distribution functions (nPDFs) \cite{Paukkunen:2017bbm}. Describing the fundamental constituents of the elements that make the world we know, nPDFs are interesting in their own right. Furthermore, they are a key input for the theoretical interpretations of a large variety of ongoing and future experiments on high-energy nuclear physics, such as heavy-ion ($A$+$A$) and proton-nucleus (p+$A$) collisions at RHIC \cite{Aschenauer:2016our} and the LHC \cite{Armesto:2015ioy, Dainese:2016gch,Salgado:2016jws}, deep-inelastic neutrino-nucleus interactions \cite{Alvarez-Ruso:2017oui} and high-energy cosmic-ray interactions in the atmosphere \cite{Bhattacharya:2016jce}. In these cases the nPDFs characterize the initial state before the collisions and, if known accurately, can lead to the discovery of 
new phenomena. Moreover, a precise knowledge of nPDFs will be crucial when searching 
for the transition between linear and non-linear scale evolution of the parton densities \cite{Albacete:2012rx,Marquet:2017bga}. 
The latter regime, known as ``saturation'' \cite{JalilianMarian:2005jf,Albacete:2014fwa}, occurs at low $x$ and low interaction scale $Q^2$ where the recombination of low-$x$ gluons becomes increasingly important.
In lepton-nucleus ($\ell$+$A$) scattering such non-linearities are predicted to be more pronounced than in 
lepton-proton ($\ell$+p) interactions \cite{Mueller:1985wy}.
Establishing non-linear effects, one of the key physics goals of an EIC, can either be done by 
comparing the behaviour of nPDFs extracted in 
different $x-Q^2$ regions with and without sizeable non-linear effects \cite{Marquet:2017bga}. 
Or in a phenomenological study of the QCD scale evolution of DIS cross sections within the framework of 
physical anomalous dimensions. There one would observe deviations from the scale evolution governed by the physical 
anomalous dimensions, which will unambiguously quantify the size and relevance of non-linear effects caused 
by an abundance of gluons with small momentum fractions \cite{Hentschinski:2013zaa}.
Altogether, the nPDFs are, and will continue to be a crucial issue in many areas of high-energy nuclear physics. 

Analogously to the free proton case, $\ell$+$A$ scattering has a huge potential to offer information on the nPDFs~\cite{Arneodo:1992wf}. Despite some considerable effort \cite{Arneodo:1996qa,Alexopoulos:2003aa}, the HERA collider was never operated with nuclear beams and thus the kinematic reach of currently available cross-section measurements in $\ell$+$A$ DIS is much more restricted than in the case of protons --- the existing fixed-target measurements do not reach $x$ much below $10^{-2}$ in the perturbative region. As a consequence, the nPDFs are significantly less constrained than the proton PDFs. 

Recently, the first global analysis of nPDFs to include LHC p+Pb Run-I data, EPPS16 \cite{Eskola:2016oht}, appeared. From the LHC data available at the time of the EPPS16 fit, the CMS dijet measurements \cite{Chatrchyan:2014hqa} had clearly the largest impact providing additional constraints on the large-$x$ gluons. Also data from electroweak boson production in p+Pb collisions were used, but their inclusion did not lead to significant improvements due to their limited statistical precision. The Run-II data with significantly higher luminosities are expected to provide much better constraints in the near future. However, theoretically robust LHC observables are limited to rather high $Q^2$ (e.g. in the case of W and Z bosons production the typical interaction scale is $Q^2 \sim 10^4 \, {\rm GeV}^2$) and it is particularly challenging to obtain reliable constraints at the low-\xs, low-$Q^2$ domain. As already mentioned, this is the important region when it comes to differentiating linear vs. non-linear scale evolution and, in general, particularly significant for bulk observables in 
heavy-ion collisions, as around 90\% of the particles produced at mid rapidity at both RHIC ($0.002 \lesssim x \lesssim 0.4$) and the LHC ($x \lesssim 10^{-3}$) come from low-$Q^{2}$ processes.

To obtain gluon constraints at small $x$ and low $Q^2$ from p+$A$ collisions at the LHC or RHIC, one has to, 
in general, rely on observables at low transverse momentum (e.g. open charm) for which theoretical 
uncertainties are significant. In order to have a cleaner probe of the partonic structure of nuclei and to extend 
the current measurements down to smaller $x$, a next-generation DIS experiment is called for. To this end, two 
possibilities have been entertained: the LHeC collider at CERN \cite{AbelleiraFernandez:2012cc} and an EIC in 
the United States \cite{Accardi:2012qut}. In the present paper, we will focus on the EIC project and its potential 
to improve the precision of nuclear PDFs. This work is organized as follows: in Sec.~\ref{sec:EIC} we present 
some technical details of an EIC, relevant for the present analysis. Secs.~\ref{sec:Observables} 
and ~\ref{sec:MonteCarlo} are dedicated to discuss the quantities that can be used to further the knowledge 
on nPDFs and showing simulation results for these, respectively. In Sec.~\ref{sec:nPDFfit} the impact of 
these measurements on the nPDFs is presented, finally in Sec.~\ref{sec:Summary} our findings are summarized.

%% file: MasterTexFiles/EIC.tex
\section{The Electron-Ion Collider project}
\label{sec:EIC}

\begin{figure}[htbp] 
   \centering
   \includegraphics[width=0.48\textwidth]{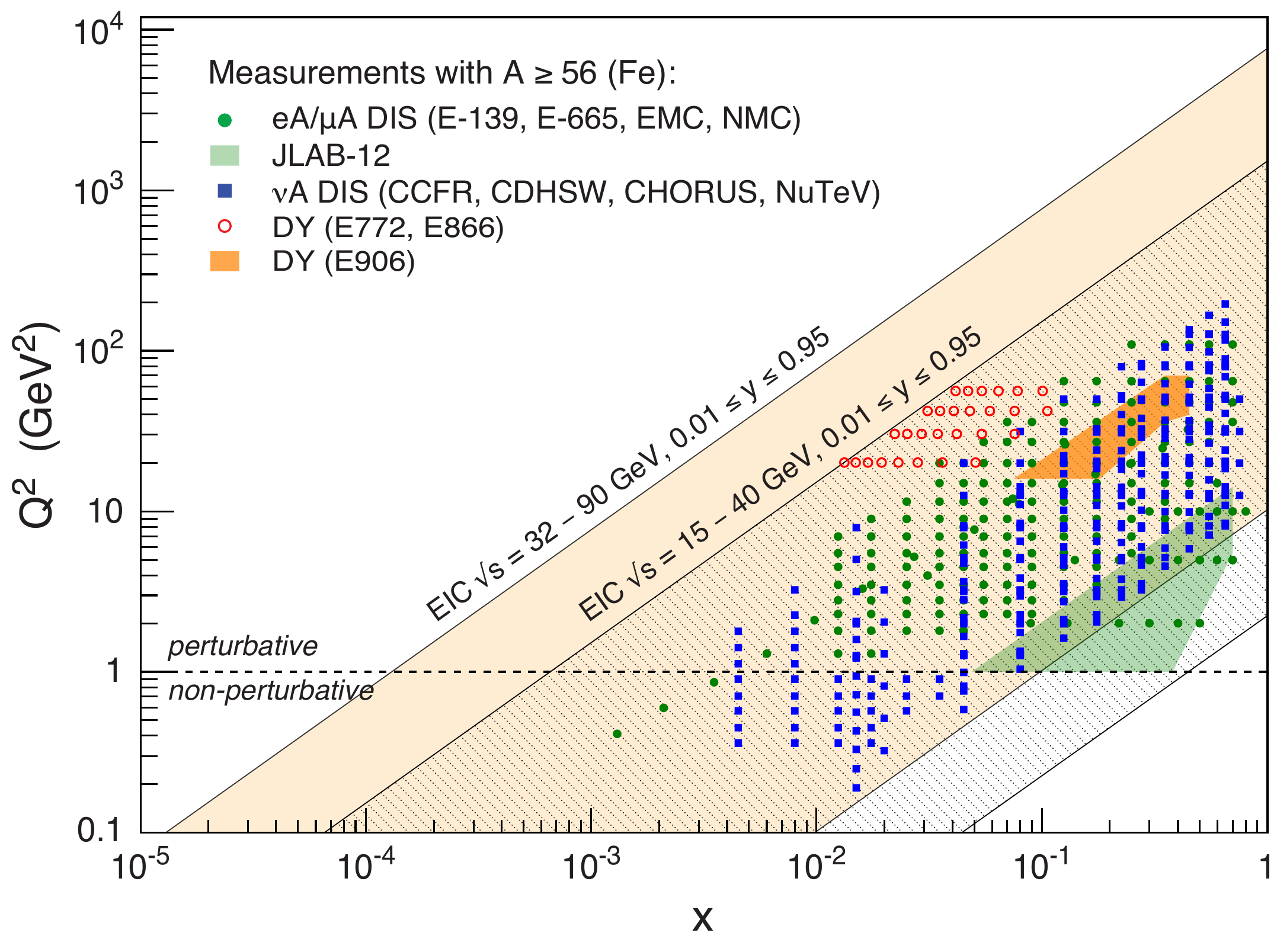}   
   \caption{The kinematic acceptance in $x$ and $Q^{2}$ of an EIC compared to completed fixed target $\ell$+$A$ DIS and Drell-Yan (DY) experiments.}
   \label{Fig:Kine-space}
\end{figure}

Currently, there are two proposals to construct an EIC in the United States. One option would involve the addition of a hadron-accelerator complex to the existing CEBAF electron facility at the Thomas Jefferson National Laboratory (JLAB), the so-called JLEIC project~\cite{Abeyratne:2015pma}. 
The other option would be to add an electron accelerator to the existing RHIC facility 
at BNL, a project know as eRHIC~\cite{Aschenauer:2014cki}.  
Despite the two proposals and strategies for an EIC, the overriding goal is the same: to build
a high-luminosity collider, which is flexible in terms of ion species (proton to uranium) and 
center-of-mass (c.o.m.) energies.
Both proposals plan for a final per-nucleon c.o.m. energies ranging from 20~GeV to 90~GeV 
for large nuclei with an even larger range (up to 145~GeV) for polarized electron+proton ($e^-$+p) collisions.
The wide kinematic coverage of an EIC, shown in Figure~\ref{Fig:Kine-space} in the \xq-plane, is very important to effectively constrain nuclear PDFs. 
Only the eRHIC proposal for an EIC could eventually be capable of reaching top c.o.m. energy at ``day 1'', 
whereas the JLEIC version would require a significant upgrade to reach the full c.o.m. energy. Therefore, JLEIC 
would stage its
measurements in c.o.m. energies, starting with scanning the high and mid $x$ region up to high $Q^{2}$ values.
Both of the proposed accelerators would also be capable to reach peak luminosities larger than 
$10^{34}$~cm$^{-2}$~s$^{-1}$, three orders of magnitude higher than what was achieved at HERA. Only the JLEIC 
version of an EIC would be capable of reaching the peak luminosity at ``day 1'', whereas eRHIC would build up 
its luminosity over time after upgrading the facility with hadron beam cooling. 
While a very large instantaneous luminosity may be required for other EIC key physics programs, this is not 
equally crucial for measuring structure functions.  As will be described later, our study proves that, assuming 
collected integrated luminosity of $10~\text{fb}^{-1}$, these measurements are - for the most part - not 
statistically limited, but rather by the associated systematic uncertainties. 
Therefore, a crucial aspect of this new accelerator complex is to match the high performance of a collider with 
a specially designed and built comprehensive DIS-specific detector in order to control systematic effects. 
The detector requirements come directly from the broad EIC science case. Some of the key capabilities such a 
detector must have are: 
\begin{itemize}
\item {\bf  Hermetic coverage in a wide pseudo-rapidity range: $\sim |\eta| \leq 4$ }
\item {\bf Good scattered lepton identification and momentum resolution:} 
in almost all cases, the DIS kinematics ($x$ and $Q^2$) of the collision are most accurately calculated from 
the scattered 
electron~\cite{Bassler:1997tv}.  Therefore, in order to measure these quantities as precisely as possible,
an excellent particle identification as well as momentum, angular resolution and  good energy resolution at 
very backward rapidities are required for the scattered lepton.

\item {\bf Good hadronic particle identification:} for semi-inclusive measurements, one is also interested in identifying the hadrons produced coincidently with the scattered lepton in the collisions. There are various 
techniques, which can be utilized to identify protons, pions and kaons at different momentum intervals.  At low 
momenta, these can be identified through their specific ionization (or dE/dx) in a time projection chamber (TPC).  
At higher momenta, Cherenkov detectors are most widely used.

\item {\bf Good secondary vertex resolution:} for measurements which involve heavy quarks (charm, bottom) a high 
resolution $\mu$-vertex  detector is essential in order to reconstruct the displaced vertices of the heavy-quark 
hadrons produced.

\item {\bf High resolution and wide acceptance forward instrumentation:} a Roman-pot spectrometer with almost
$100\%$ acceptance and a wide coverage in scattered proton four-momentum is crucial for studies of diffractive 
physics in $e^-$+p and $e^-$+A collisions. Furthermore, for $e^-$+A collisions, a zero-degree calorimeter (ZDC) 
with sufficient acceptance is a key feature vetoing on the nucleus break-up and determining the impact 
parameter of the collision~\cite{Zheng:2014cha}.
\end{itemize}

%% file: MasterTexFiles/Observables.tex
\section{Reduced cross section and longitudinal structure function}
\label{sec:Observables}

\begin{figure}[htbp]
	\vspace{0mm}
	\begin{center}
	   \includegraphics[width=0.45\linewidth]{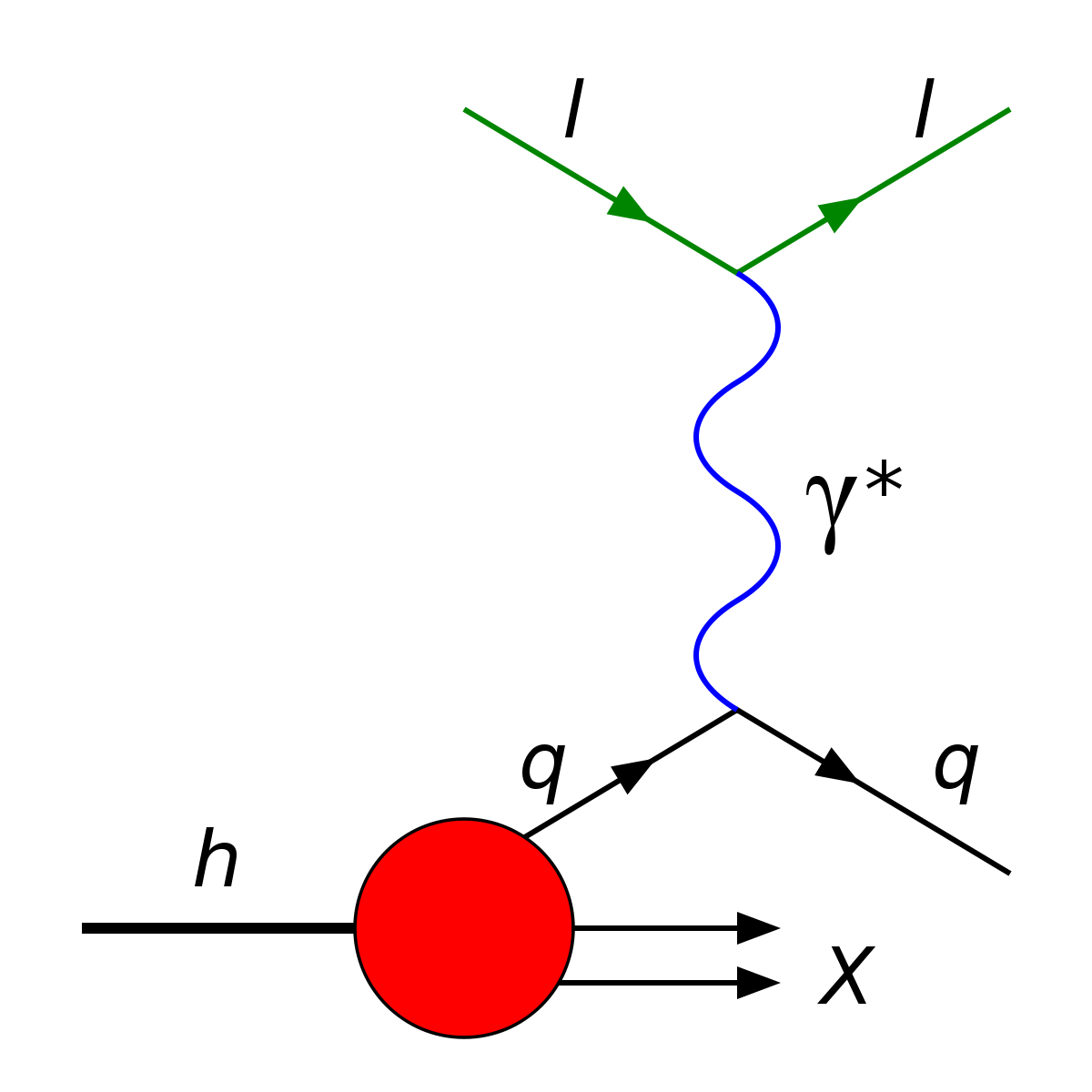} 
		\includegraphics[width=0.45\linewidth]{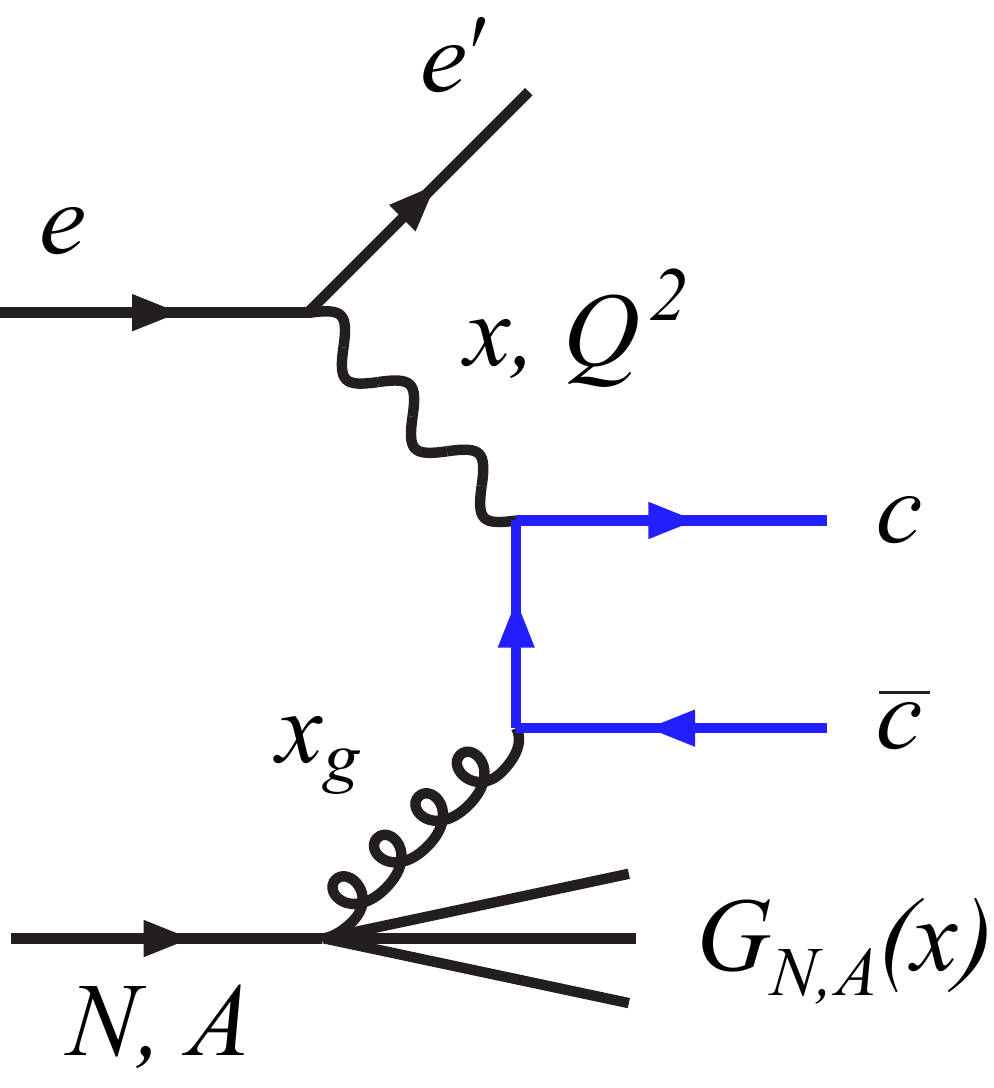} 
	\end{center}
	\caption{\emph{Left}: A depiction of inclusive DIS. \emph{Right}: $c\bar{c}$ production through photon-gluon fusion.}
	\label{fig:DIS} 
\end{figure}

The inclusive DIS process is a hard interaction between a lepton and a nucleon, in which the latter breaks up, the invariant mass of the hadronic final state being much larger than the nucleon mass. This is depicted in the left diagram of Figure \ref{fig:DIS}.
All the relevant kinematic variables that describe the interaction are defined in Table~\ref{Tab:kin}. 

\begin{table}[htbp!]
\begin{center}
\caption {Relevant kinematical variables in a DIS process.} \label{Tab:kin} 
  \begin{tabular}{cl}
    {\bf Variable} & {\bf Description} 
    \tabularnewline \hline
    $\eta$ & pseudo-rapidity of particle    
    \tabularnewline   
    $x$ & fraction of the nucleon momentum 
    \tabularnewline   
        &  carried by the struck parton 
    \tabularnewline
    $y$ & inelasticity, fraction of the lepton's energy lost 
    \tabularnewline
       & in the nucleon rest frame. 
    \tabularnewline
    $\sqrt{s}$ & center-of-mass energy 
    \tabularnewline
    $Q^2$ & squared momentum transferred to the lepton, 
    \tabularnewline
          & equal to the virtuality of the exchanged photon 
    \tabularnewline
          & Note the relation $Q^2 \approx x y s$.
\end{tabular}
 \end{center} 
\end{table}

The direct observable used for constraining the nPDF is the cross section ($\sigma$), which is customarily expressed 
as a dimensionless quantity known as ``reduced'' cross section
$\sigma_{\rm r}$, defined as
\begin{equation}
\sigma_{\rm r} \equiv
\left(\frac{d^2\sigma}{dxdQ^2}\right)\frac{xQ^4}{2\pi\alpha_{\rm em}^2[1+(1-y)^2]},
\label{Eq:RedCrossSec}
\end{equation}
where $\alpha_{\rm em}$ is the QED fine-structure constant. At small $x$, the reduced cross section can be approximately expressed in terms of the structure function $F_2$ and the longitudinal structure function $F_{\rm L}$ as
 
\begin{equation}
\sigma_{\rm r} = 
                  F_2(x,Q^2) - \frac{y^2}{1+(1-y)^2}F_{L}(x,Q^2).
\label{Eq:StructFunc}
\end{equation}

\begin{figure*}[htbp!] 
   \centering
   \includegraphics[width=0.3\textwidth]{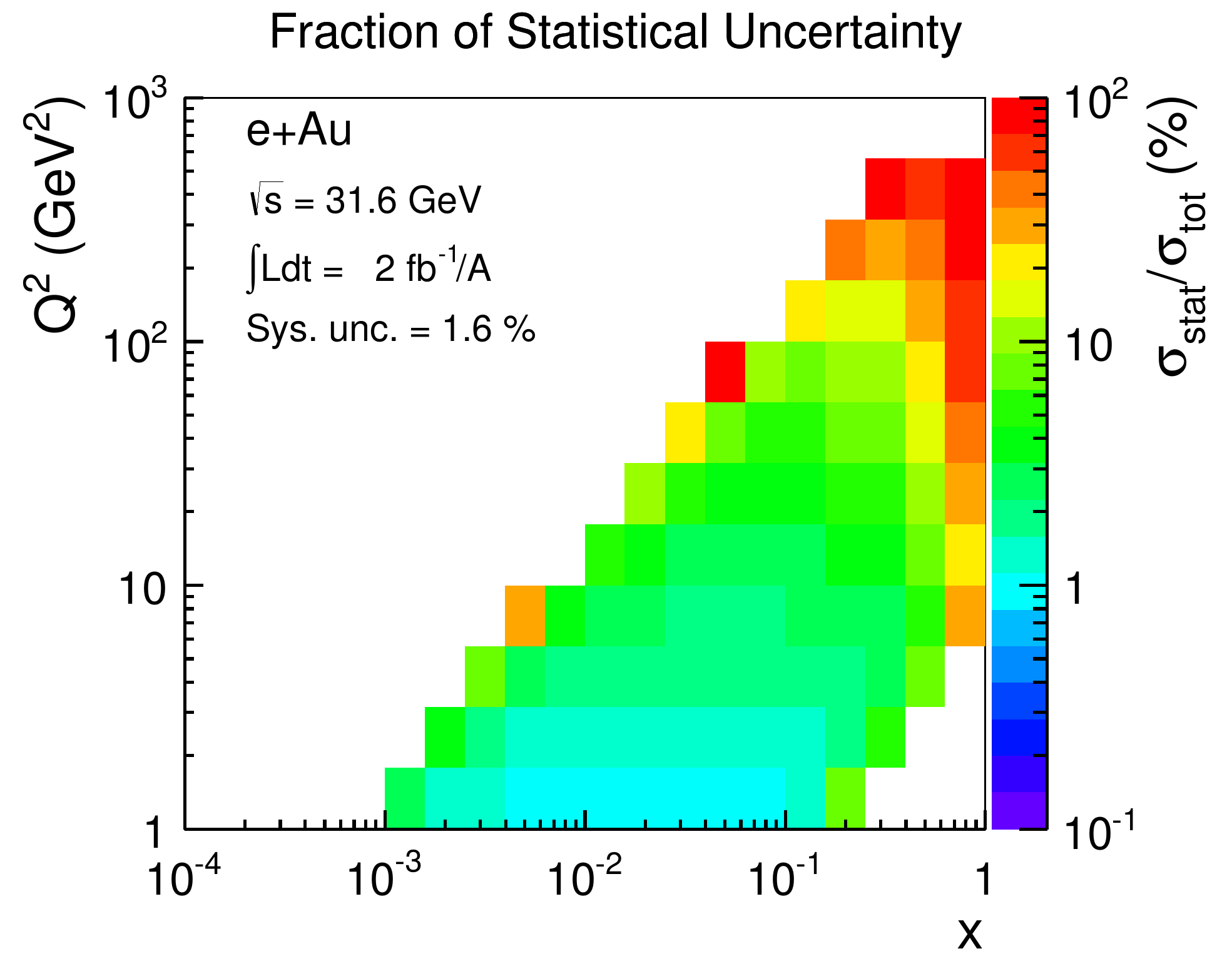}   
   \includegraphics[width=0.3\textwidth]{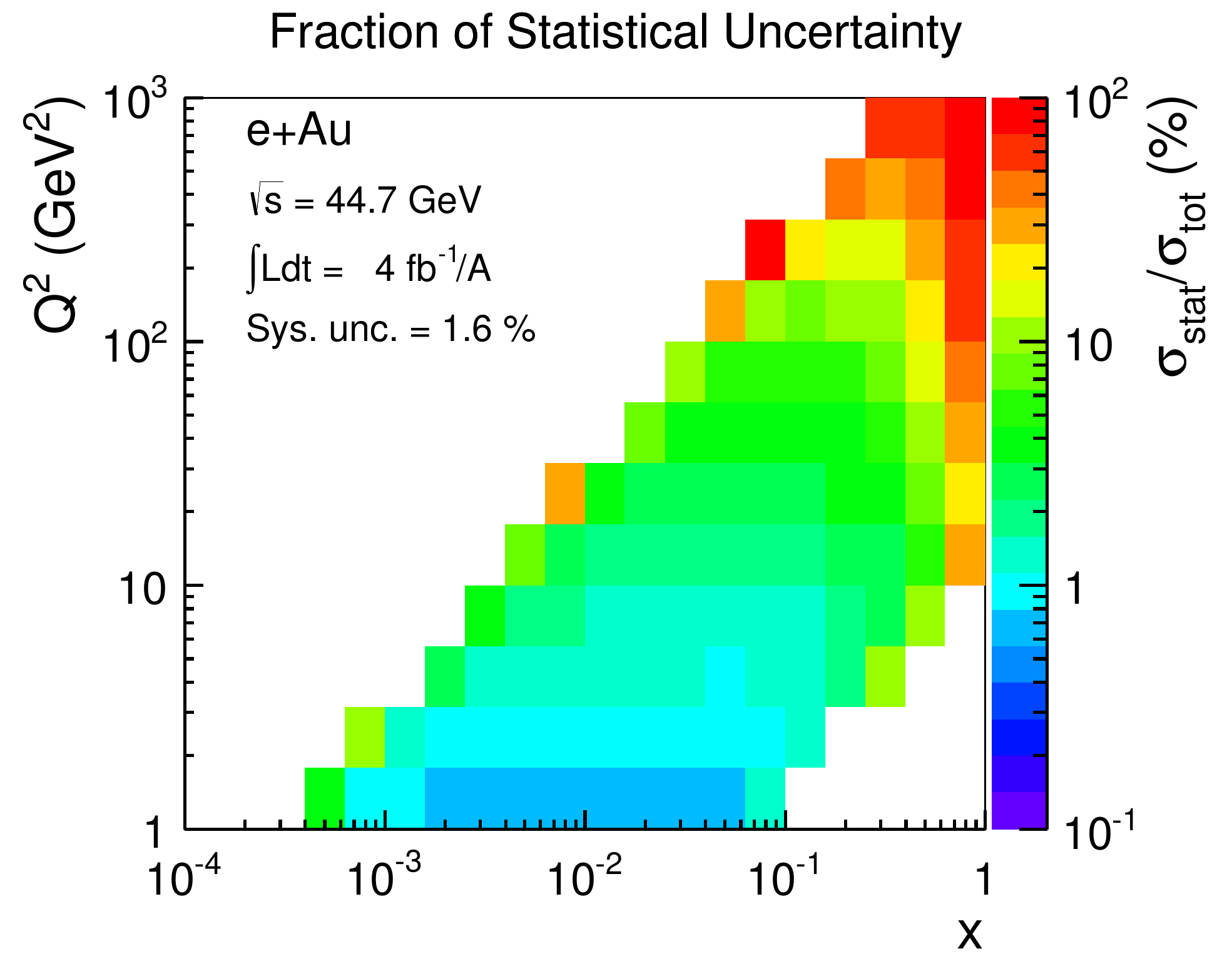} 
   \includegraphics[width=0.3\textwidth]{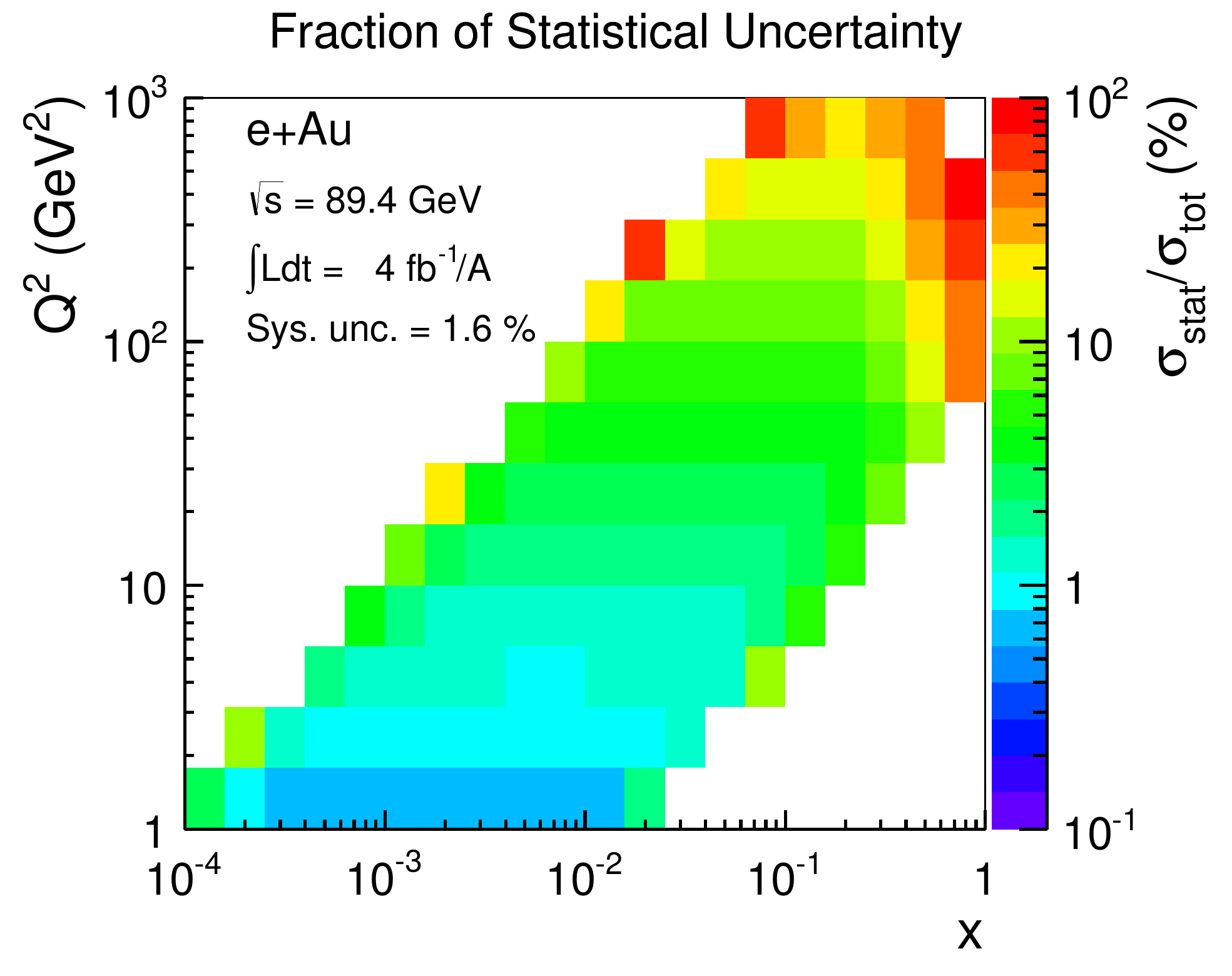} 
   \caption{Fraction of statistical uncertainty over the total uncertainty of the simulated 
reduced cross section measurement at an EIC, for each $x$ and $Q^{2}$ bin, at different c.o.m. 
energies, assuming a combined collected luminosity of 10~fb$^{-1}$.}
  \label{Fig:uncertainty}   
\end{figure*}

While $F_{2}$ is sensitive to the momentum distributions of (anti)quarks, and to gluons mainly through scaling 
violations, $F_{\rm L}$ has a larger direct contribution from gluons \cite{Armesto:2010tg}.
In most of the kinematical space covered by the old fixed-target DIS experiments, $\sigma_{\rm r}$ is dominated 
by $F_2$, to the extent that the older data were presented solely in terms of $F_2$, 
largely disregarding $F_{\rm L}$. 
Therefore the information on $F_{\rm L}$ and, consequently, the direct access to the nuclear gluon are not 
currently available. At an EIC, the high luminosity and wide kinematic reach will enable the direct extraction 
of $F_{\rm L}$ and thereby more information on the behaviour of the nuclear gluons can be obtained.
In addition, an EIC will offer possibilities to constrain the gluon density in nuclei via 
measurements of the charm (bottom) structure function on which only one prior measurement exists 
\cite{Aubert:1982tt}. Heavy quarks, due to their large mass, are mainly produced through photon-gluon fusion 
(as illustrated in the right diagram of Figure \ref{fig:DIS}), the measurement of the corresponding reduced cross 
section $\sigma_{\rm r}^{c\bar{c}}$ provides complementary information on the gluon distribution in nuclei. Also, the so far unmeasured charm contribution to $F_{\rm L}$ will be measurable at an EIC. 

In the production of heavy quarks, the effects of quark mass $m_{q}$ require a careful treatment to preserve the genuine, dynamical effects of $m_{q}$ in the partonic processes at low-$Q^2$ region ($Q^2 \lesssim m_{q}^2$), but also to have a well defined asymptotic limit ($Q^2 \gg m_{q}^2$). This has lead to the development of the so-called general-mass variable flavor number scheme (GM-VFNS) which is nowadays routinely implemented in proton- and nuclear-PDF extractions. The implementation of GM-VFNS is not unambiguous, but inherently contains certain scheme dependence and several 
versions of the GM-VFNS can be found in the literature, see. e.g. Ref. \cite{Thorne:2008xf}. 
Furthermore the theory can be formulated in terms of the running or the pole mass \cite{Alekhin:2010sv,Accardi:2016ndt}.
The possibility of a precise measurement of heavy-flavour observables at an EIC, in particular the so far unmeasured 
charm contribution to $F_{L}$, will offer an opportunity to benchmark different schemes with an unprecedented precision. In addition, an EIC will take the possibilities to constrain the intrinsic heavy-flavour components in PDFs onto a completely new level.


In Table~\ref{Tab:nPDFs}, we summarize some properties of the observables we have discussed. For the reduced cross sections the kinematic reach is always wide and they can be measured in practically everywhere within the regions indicated in Figure~\ref{Fig:Kine-space}. In the case of longitudinal structure functions the kinematic range is more restricted as their extractions require measurements at fixed $x$ and $Q^2$ with several c.o.m. energies. However, e.g. the smallest values of $x$ can only be reached at the top c.o.m. energy and thus no $F_{\rm L}$ measurement can be performed there. This will be further discussed in the next section. Also, the sensitivity to the gluon PDFs is indicated.

\begin{table*}[htbp!]
\begin{center}
\caption {Properties of the observables.} \label{Tab:nPDFs}
  \begin{tabular}{|c||c|c|c|c|}\hline
 & $\sigma_{\rm r}$ & $F_{\rm L}$ & $\sigma_{\rm r}^{c\bar{c}}$ & $F_{\rm L}^{c\bar{c}}$ \\ \hline \hline
 Kinematic coverage & wide & limited & wide & limited \\ \hline
 Access to gluons & mainly via scale evolution & direct & direct &  direct \\ \hline
  \end{tabular}
 \end{center} 
\end{table*}

%% file: MasterTexFiles/MonteCarlo.tex
\section{Monte Carlo simulations}
\label{sec:MonteCarlo}

\subsection{Inclusive reduced cross sections}

To estimate the statistical uncertainties in measuring $\sigma_{\rm r}$ in $e^-$+$A$ collisions, we simulated events using the PYTHIA 6.4~\cite{Sjostrand:2006za} Monte Carlo (MC) generator with 
EPS09~\cite{Eskola:2009uj} nuclear PDFs, for different beam-energy configurations corresponding to 
a range in c.o.m. energy from 30 to 90~GeV. 
We assumed the following c.o.m. energies: $\sqrt{s}= 31.6, 44.7$, and 89.4~GeV. 
In doing so we simulated a data collection of 2~fb$^{-1}$ integrated luminosity at 
$\sqrt{s}= 31.6$~GeV and to 4~fb$^{-1}$ at $\sqrt{s}= 44.7, 89.4$~GeV respectively, corresponding 
to a combined 10~fb$^{-1}$. We divided our phase space in $5 \times 4$ bins per decade in 
$x$ and $Q^{2}$. For the purpose of this study we conservatively assumed a bin-by-bin systematic 
uncertainty of $1.6\%$ based on what has been achieved at HERA.
We also consider an additional overall $1.4\%$ systematic uncertainty originating from the 
luminosity measurement. Figure~\ref{Fig:uncertainty}  shows the fraction of the statistical 
uncertainty over the total one (with systematics added in quadrature), per each bin in 
$x$ and $Q^{2}$ for $\sigma_{\rm r}$. 
One can see that the $\sigma_{\rm r}$ determintion is generally dominated by the systematic uncertainties. 
Nevertheless, reducing the statistical uncertainty may become relevant when extending the 
investigation to high values of $x$ at very high $Q^{2}$, where collecting a data sample of 
a significantly higher integrated luminosity may be required for precision measurements.

\begin{figure*}[htbp] 
   \centering
   \includegraphics[width=0.45\textwidth]{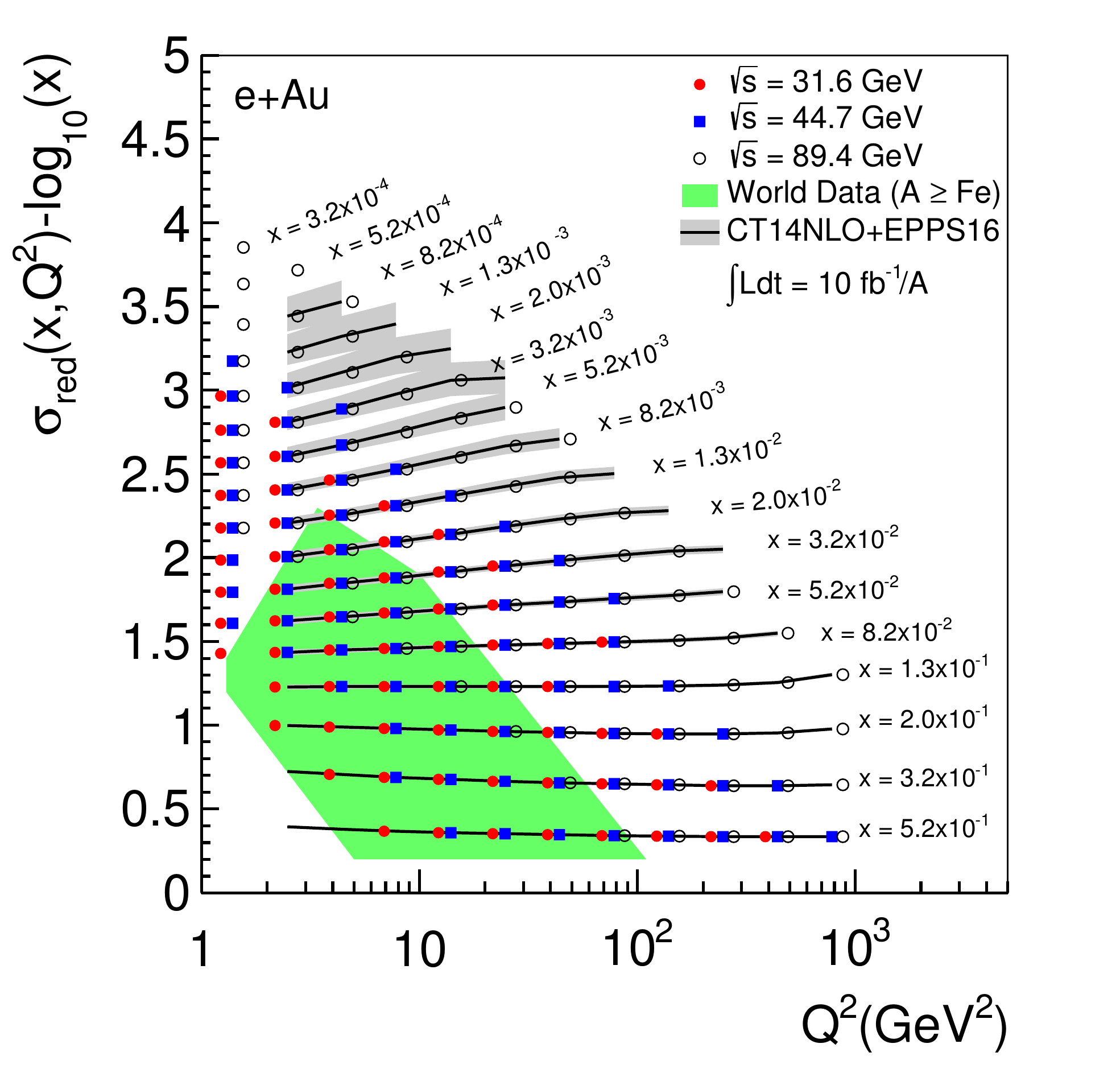}   
   \includegraphics[width=0.45\textwidth]{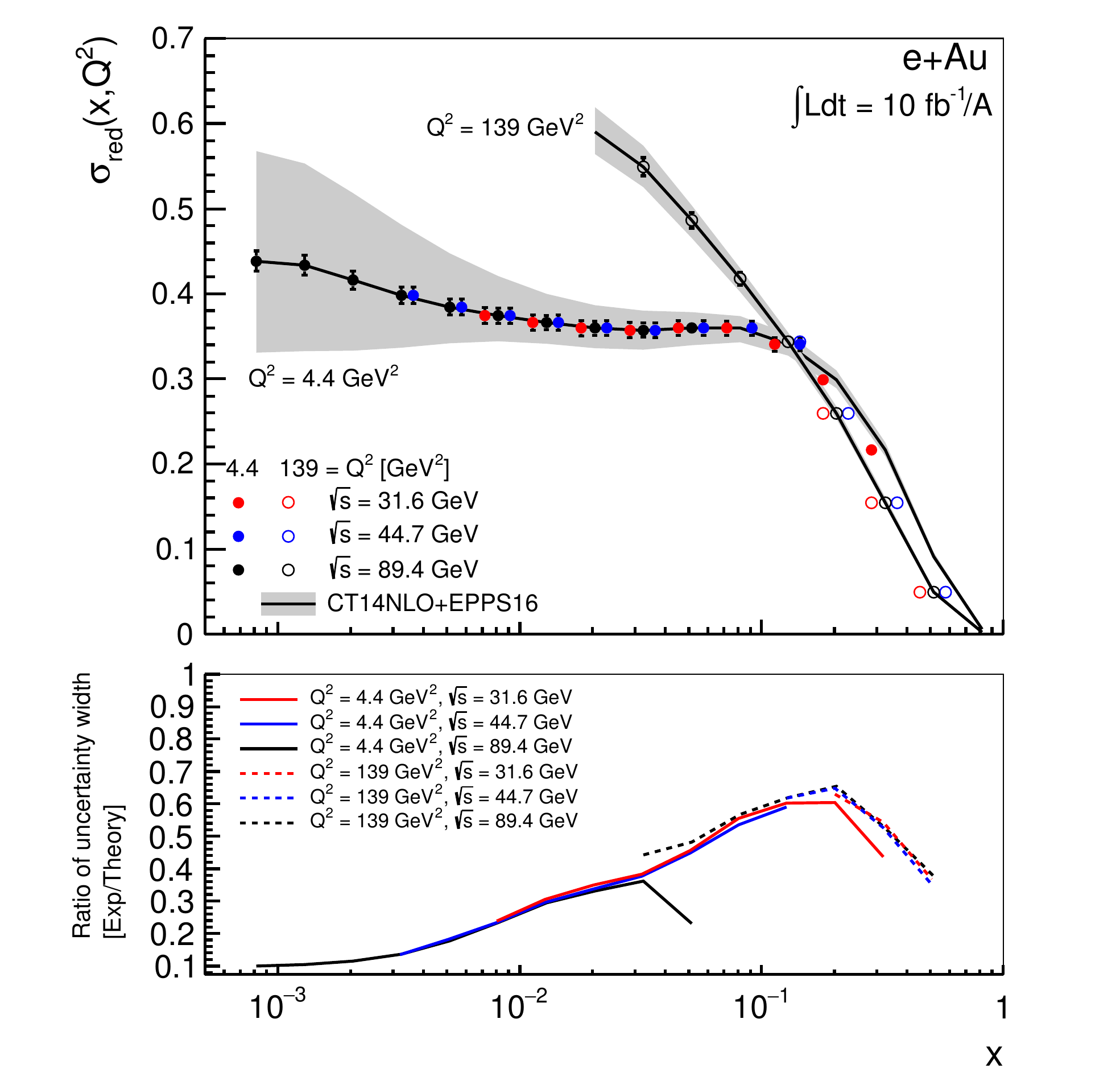}
   \caption{The reduced cross section ({\it left}) in $e$+Au collisions at EIC is plotted as a 
function of $Q^2$ and $x$, the kinematic space covered by currently available experimental data 
is marked on the plot by the the green area. The measured reduced cross section points are shifted 
by $-\text{log}_{10}(x)$ for visibility. Two examples of the $\sigma_{r}$ ({\it right}) at $Q^{2}$ 
values of 4.4~GeV$^{2}$ and 139~GeV$^{2}$  are plotted versus $x$, 
with the ratio between the widths of the experimental and theoretical uncertainties shown in the bottom panel.
In both plots the statistical 
and systematic uncertainties are added in quadrature and compared to the theory uncertainty 
({\it gray bands}) from CT14NLO+EPPS16. The overall $1.4\%$ systematic uncertainty on the 
luminosity determination in not shown on the plots. Points that correspond to different energy 
configurations are horizontally offset in $Q^2$ for visibility.}
   \label{Fig:SigmaRed-F2}
\end{figure*}

Figure \ref{Fig:SigmaRed-F2}~({\it left}) shows $\sigma_{\rm r}$ for $e^-$+Au collisions plotted versus 
$Q^2$ at different $x$ values, for the three c.o.m. energies. A comparable precision can be 
achieved using any other nucleus in a similar kinematical range.  The current experimental DIS-data
coverage for large nuclei (A $\geq$ Fe) is also shown and, for clarity, $\sigma_{\rm r}$ is offset by 
subtracting log$_{10}$($x$) and points corresponding to different energies are horizontally offset in $Q^2$. 
The bin-by-bin statistical and systematical uncertainties are added in quadrature, whereas the 
overall systematic uncertainty of 1.4\% on the luminosity determination is not shown. 
The central values for the data points have been adjusted to a next-to-leading order (NLO) calculations with 
CT14NLO \cite{Dulat:2015mca} free proton PDFs supplemented with the latest nuclear modifications from 
EPPS16~\cite{Eskola:2016oht}.

Two examples of the $\sigma_{\rm r}$ as a function of $x$ at $Q^{2} = 4.4$ and 139~GeV$^{2}$ are shown in Figure 
\ref{Fig:SigmaRed-F2}~({\it right}) and compared with the theory uncertainties from EPPS16 and CT14NLO. 
The bottom panel shows the ratios between the full widths of the experimental and theoretical uncertainties 
versus $x$ for the different c.o.m energies.
At small $x$, and small $Q^2$ in particular, the expected uncertainties on inclusive cross-section measurements 
at an EIC are much smaller than those from the prediction based on EPPS16 and CT14NLO (grey band). Towards larger 
values of $x$, the existing constraints from old fixed-target experiments (SLAC and NMC in particular) do already 
provide stringent constraints for nPDFs and thus the advantage of EIC measurements on $\sigma_{\rm r}$ lies 
predominantly at small $x$. The estimated impact that these inclusive EIC data will have on the current 
knowledge of nuclear PDFs will be discussed in Section~\ref{sec:nPDFfit}.

\subsection{Reduced cross section in charm production}

\begin{figure*}[htbp] 
   \centering
      \includegraphics[width=0.40\textwidth]{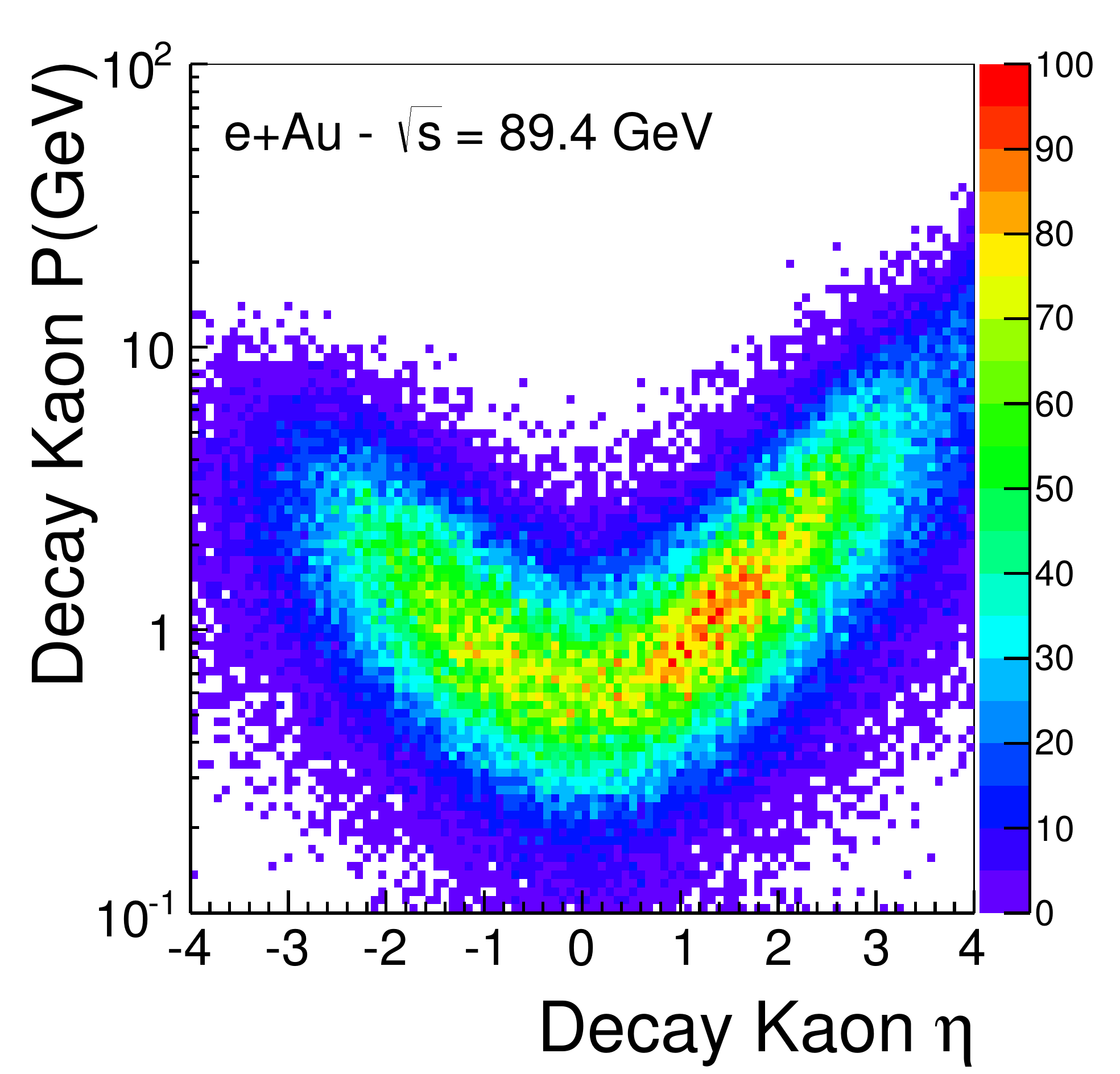} 
      \includegraphics[width=0.45\textwidth]{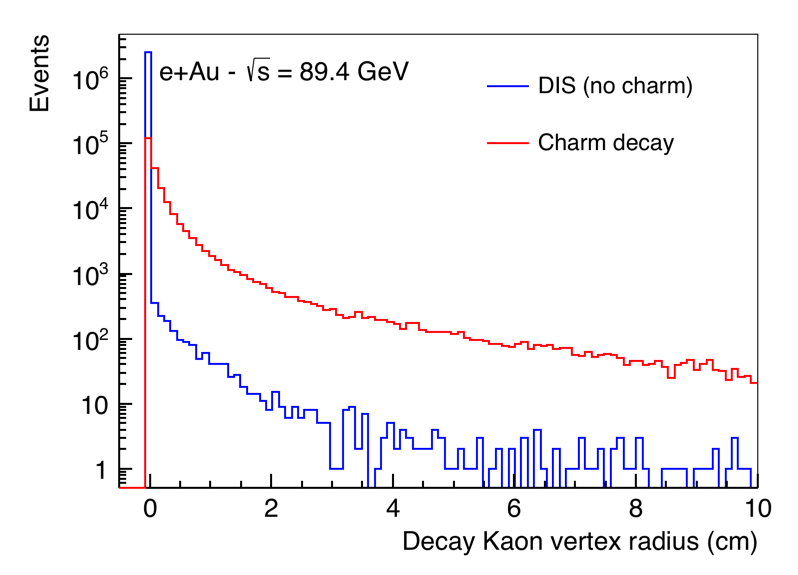} 
   \caption{Left: The distribution of the momentum of a decay $K$ from $c\bar{c}$ production 
events versus pseudo-rapidity. Right: The vertex position of $K$ in inclusive DIS ({\it blue line}) 
compared to $c\bar{c}$ production events ({\it red line}).}
   \label{Fig:KaonsKine}
\end{figure*}

Within the simulated data sample we have also selected $c\bar{c}$ production events by tagging 
$K$ mesons which are decay products of the $D$ mesons produced in the charm fragmentation. 
Figure~\ref{Fig:KaonsKine} shows the momentum  distribution of the decay kaons in a charm production events as a
function of  pseudo-rapidity ({\it top plot}) and the 
distribution of the vertex position of kaons in inclusive DIS compared with charm production 
events ({\it bottom plot}). One can see that kaons with a displaced vertex are coming predominently 
from $c\bar{c}$ decay and are mainly produced at $\eta \leq |3| $ with momenta below 10~GeV.

Based on the vertex distribution in Figure~\ref{Fig:KaonsKine} to suppress the background from 
non-charm events we have requested the $K$ to come from a vertex displaced between 0.01 and 3~cm 
with respect to the interaction point. 
Additional selection requirements on the $K$ momentum ($p_{K}$), have been imposed to account for 
the $\eta$-acceptance of the particle identification (PID) detectors integrated in the EIC detector 
as shortly described in Sec.~\ref{sec:EIC}. 
We have assumed the following $K$ PID technologies to be at place: 
At mid-rapidity ($-1 < \eta < 1$), energy loss ($dE/dx$) in the central tracker (i.e. a time-projection chamber), 
and a proximity focusing Aerogel Ring-Imaging Cherenkov (RICH) detector covering the $K$ momentum ranges 
$0.2~\text{GeV} < p_{K} < 0.8~\text{GeV}$ and $2~\text{GeV} < p_{K} < 5~\text{GeV}$, respectively.
We considered at forward rapidities ($1 < \eta < 3.5$) a dual radiator RICH covering the kaon 
momentum range $2~\text{GeV} < p_{K} < 40~\text{GeV}$, and at backward rapidities 
($-3.5 < \eta < -1$) an Aerogel RICH covering $2~\text{GeV} < p_{K} < 15~\text{GeV}$.  

\begin{figure*}[htbp!] 
   \centering
      \includegraphics[width=0.48\textwidth]{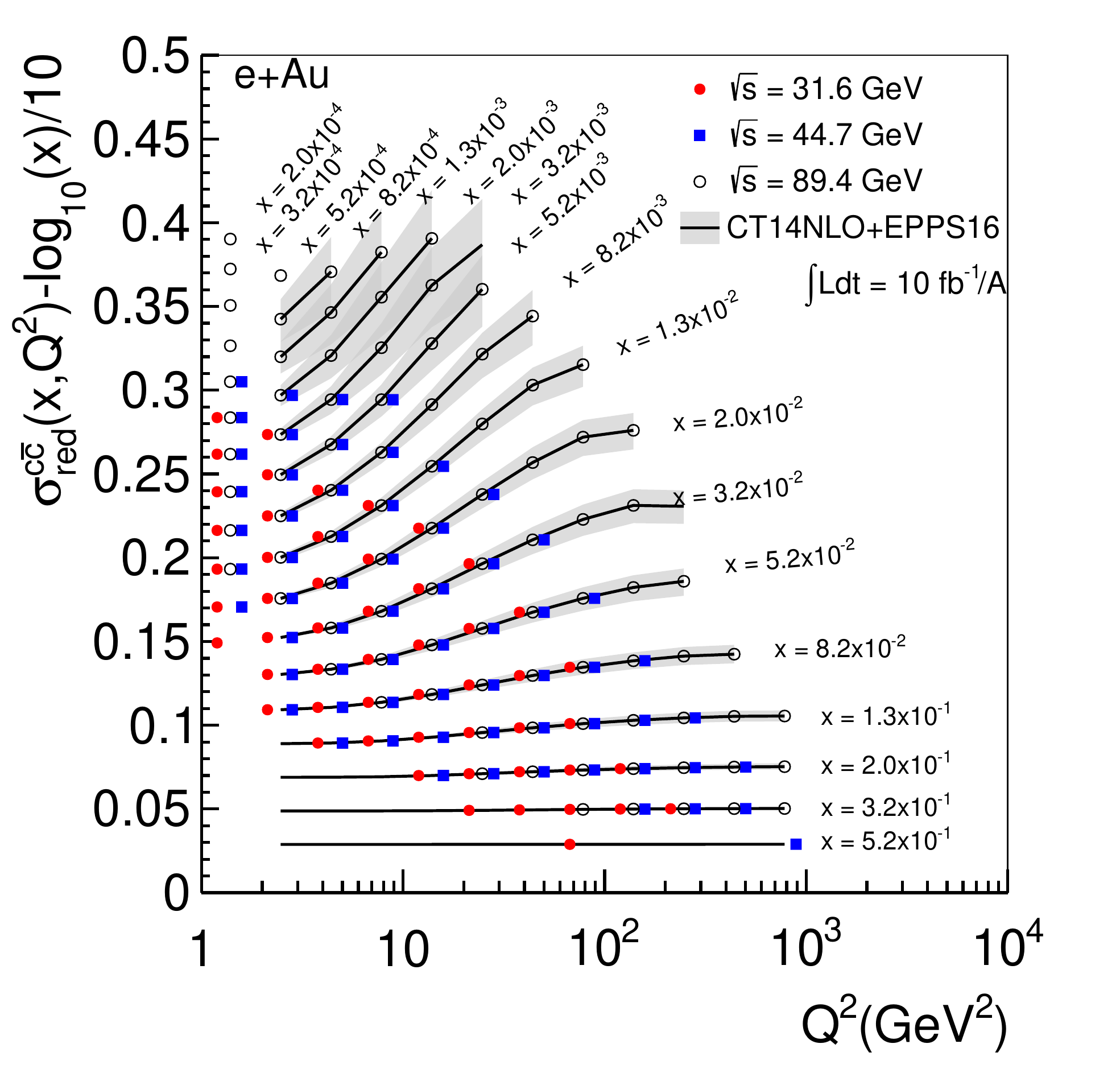}   
      \includegraphics[width=0.48\textwidth]{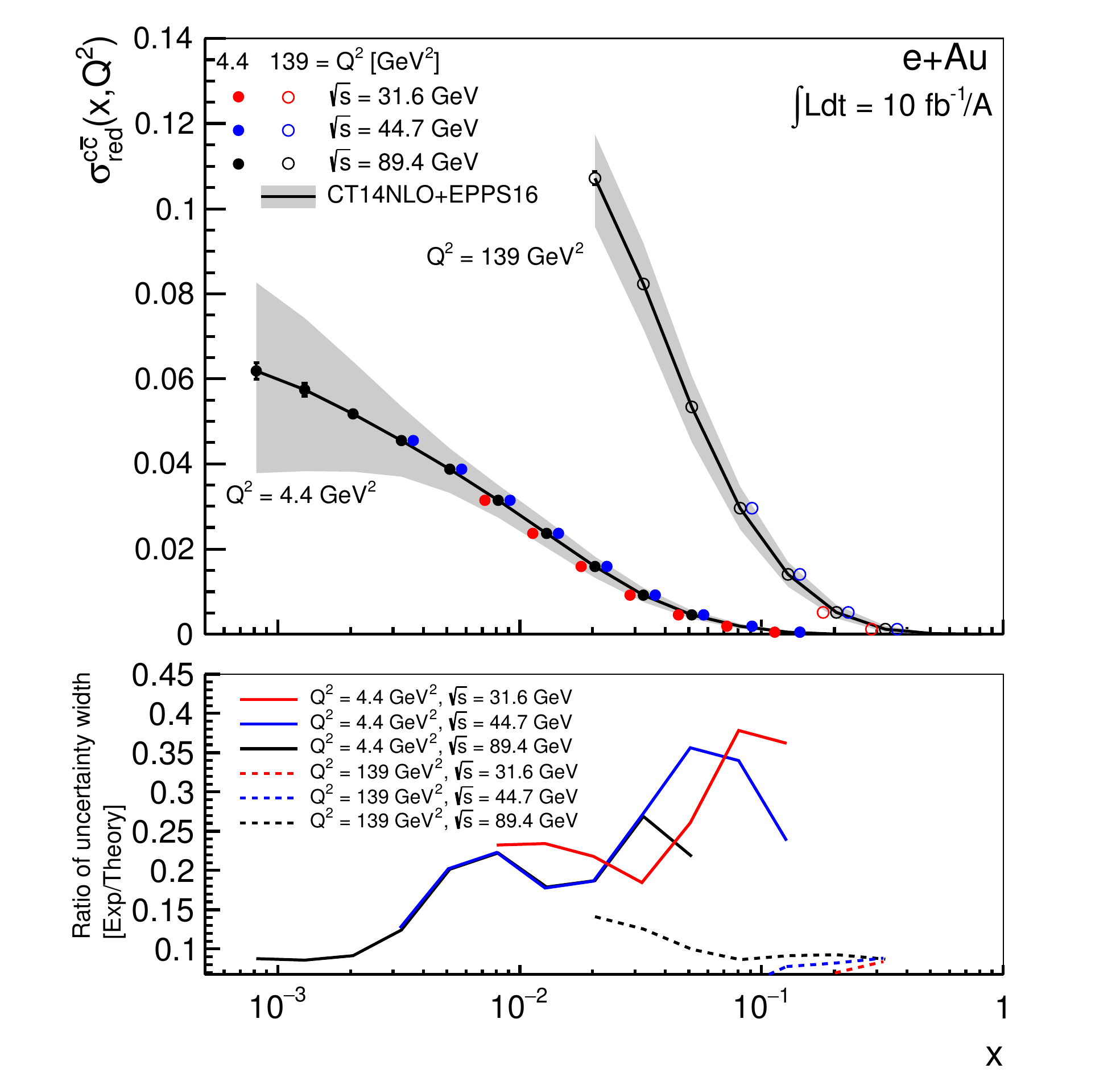}
   \caption{The reduced cross section ({\it left}) of $c\bar{c}$ production in $e$+Au collisions at an EIC is 
plotted as a function of $Q^2$ and $x$. The points are shifted by $-\text{log}_{10}(x)/10$ for visibility. Two 
examples of the $\sigma_{r}^{c\bar{c}}$ ({\it right}) at $Q^{2}$ values of 4.4~GeV$^{2}$ and 139~GeV$^{2}$ are 
plotted versus $x$, 
with the ratio between the widths of the experimental and theoretical uncertainties shown in the bottom panel.
In both plots the statistical and systematic uncertainties are added in quadrature and compared 
to the theory uncertainty ({\it gray bands}) from CT14NLO+EPPS16. The overall $1.4\%$ systematic uncertainty on 
the luminosity measurement in not shown on the plots. Points that correspond to different energy configurations 
are horizontally offset in $Q^2$ for visibility.}
\label{Fig:F2c}
\end{figure*}

The assumed bin-by-bin systematic uncertainty for the measurement of $\sigma^{c\bar{c}}_{\rm r}$ 
is 3.5\% 
and it is added in quadrature to the statistical one. The point to point systematics is 
higher to account for the additional challenge to positively identify the kaon in the particle ID 
detectors. An overall 1.4\% systematic uncertainty originating from the measurement of luminosity 
is also assumed. 

Figure \ref{Fig:F2c}~({\it left}) shows $\sigma^{c\bar{c}}_{\rm r}$ plotted versus $Q^2$ at different $x$ values, 
for the selected c.o.m. energies used earlier. For clarity, $\sigma^{c\bar{c}}_{\rm r}$ is offset by subtracting 
$\text{log}_{10}(x)/10$ and points that correspond to different energy configurations are horizontally offset in 
$Q^2$. 
Figure~\ref{Fig:F2c}~({\it right}) also shows two examples of the $\sigma_{r}^{c\bar{c}}$ as a function of $x$ at 
$Q^{2 } = 4.4$ and 139~GeV$^{2}$.
The bottom panel shows the ratios between the full widths of the experimental and theoretical uncertainties versus $x$ 
for the different c.o.m energies.
The assumed overall uncertainty on the luminosity is not shown on the plots. As done for the inclusive study, 
also for the charm production the data points have been rescaled onto the predictions from CT14NLO+EPPS16. 
The expected uncertainties on $\sigma_{\rm r}^{c\bar{c}}$ at an EIC are much smaller than the prediction based 
on EPPS16 (grey band). Unlike in the inclusive case, for charm production the theory uncertainties clearly exceed 
the projected experimental ones also at large $x$.

The efficiency of selecting $c\bar{c}$ production events has been evaluated as the ratio between the number of 
selected charm events and the number of all charm events simulated within the kinematical acceptance of an EIC. 
The overall charm selection efficiency has been estimated to be $\sim30\%$ with no significant c.o.m. energy 
dependence.
A slight rise with $x$ was also found, but it is not significant at very small $Q^{2}$ values and becomes a 
little more pronounced at higher $Q^{2}$. 

In order to be confident that the selection criteria used in the present study yield a sufficiently clean sample 
of charm production events, we studied possible background contaminations. The ratio between the number of 
background events with kaons in the final state passing the whole selection but not coming from a charm decay, 
and the signal containing only charm events has been studied. The overall background over signal ratio (B/S) has 
been estimated to be respectively 0.95\% ($\sqrt{s} = 31.6$~GeV), 0.98\% ($\sqrt{s} = 44.7$~GeV), and 1.16\%  
($\sqrt{s} = 89.4$~GeV), thus showing a slight c.o.m. energy dependence.
B/S has been also studied as a function of $x$ at different $Q^{2}$ values for the selected energies and it 
was found to never significantly exceed 2\%.

\subsection{QED Corrections }

Cross section measuremeants with a precission as anticipated from an EIC need to account for all processes, 
which could alter the relation of measured to true event kinematics.    
The radiation of photons and the corresponding virtual corrections (QED corrections) from the incoming and 
outgoing lepton can cause significant effects on the reconstruction of the reduced cross-section. 
The correction of these radiative effects can be either done through Monte-Carlo techniques or 
including the QED effects directly in the PDF analysis. 

For neutral-current $l+A$ scattering, there exists a
gauge-invariant classification into leptonic, hadronic and interference contributions. 
The dominant correction comes from the leptonic contribution, where the photons are emitted collinear 
with the leptons and give rise to large logarithmic terms 
$\propto \log(Q^2/m_{\rm \ell}^2)$, where $m_{\rm \ell}$ is the lepton mass. In comparison to the case 
with no radiation, the momentum carried by the radiated photons will alter the values of $x$ and $Q^2$ 
measured from the scattered lepton. Since the PDFs are typically very steep functions of $x$, even small 
changes can lead to large variation in the cross sections. Also the initial- and final-state quarks may 
radiate photons giving rise to large logarithmic terms, which are nowadays often resummed to photonic 
component in the PDFs. However, these corrections do not alter the event kinematics and are therefore 
much smaller than the contributions coming from the radiation off the leptons.

\begin{figure}[htbp] 
   \centering
   \includegraphics[width=0.23\textwidth]{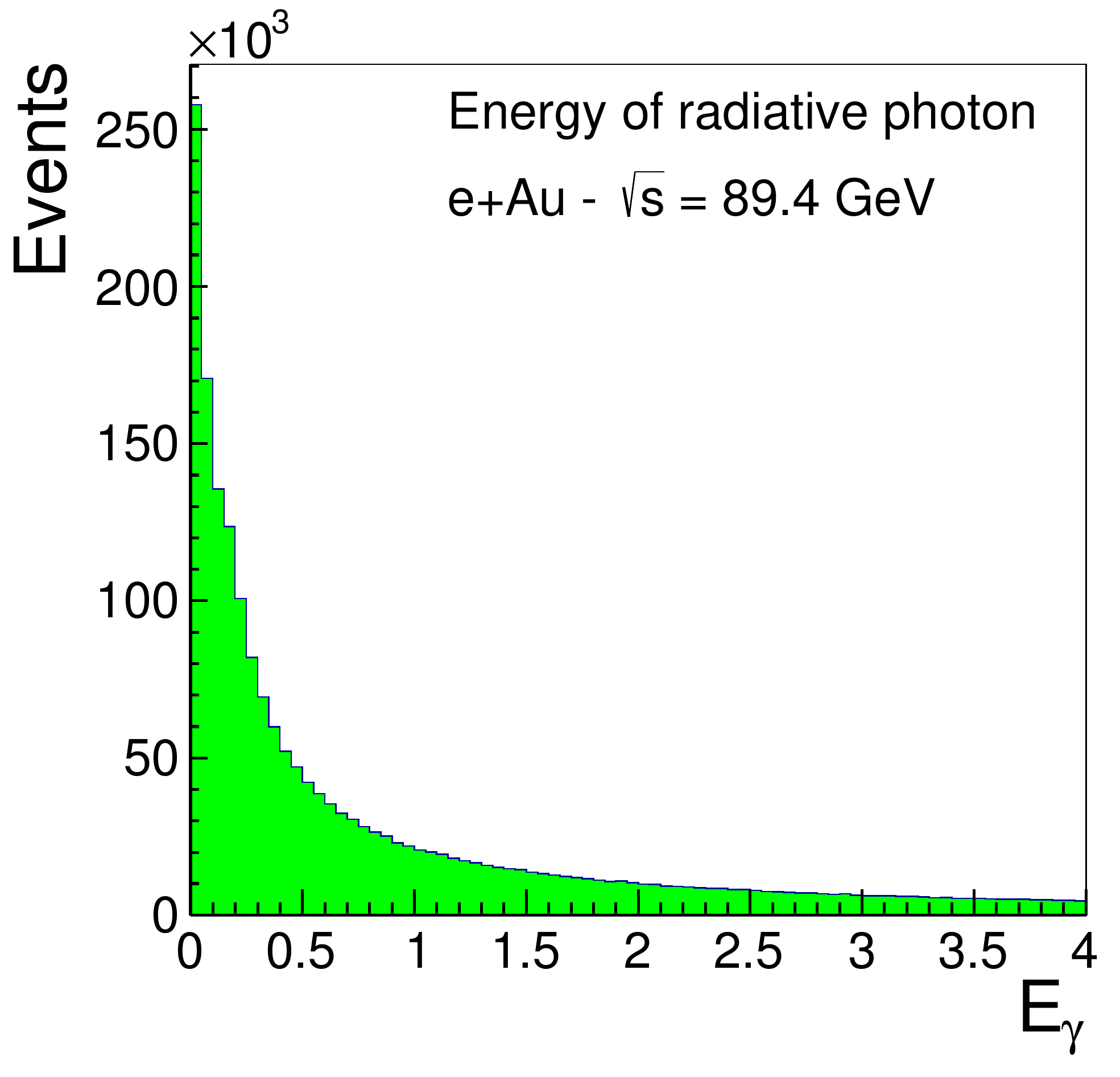}
   \includegraphics[width=0.23\textwidth]{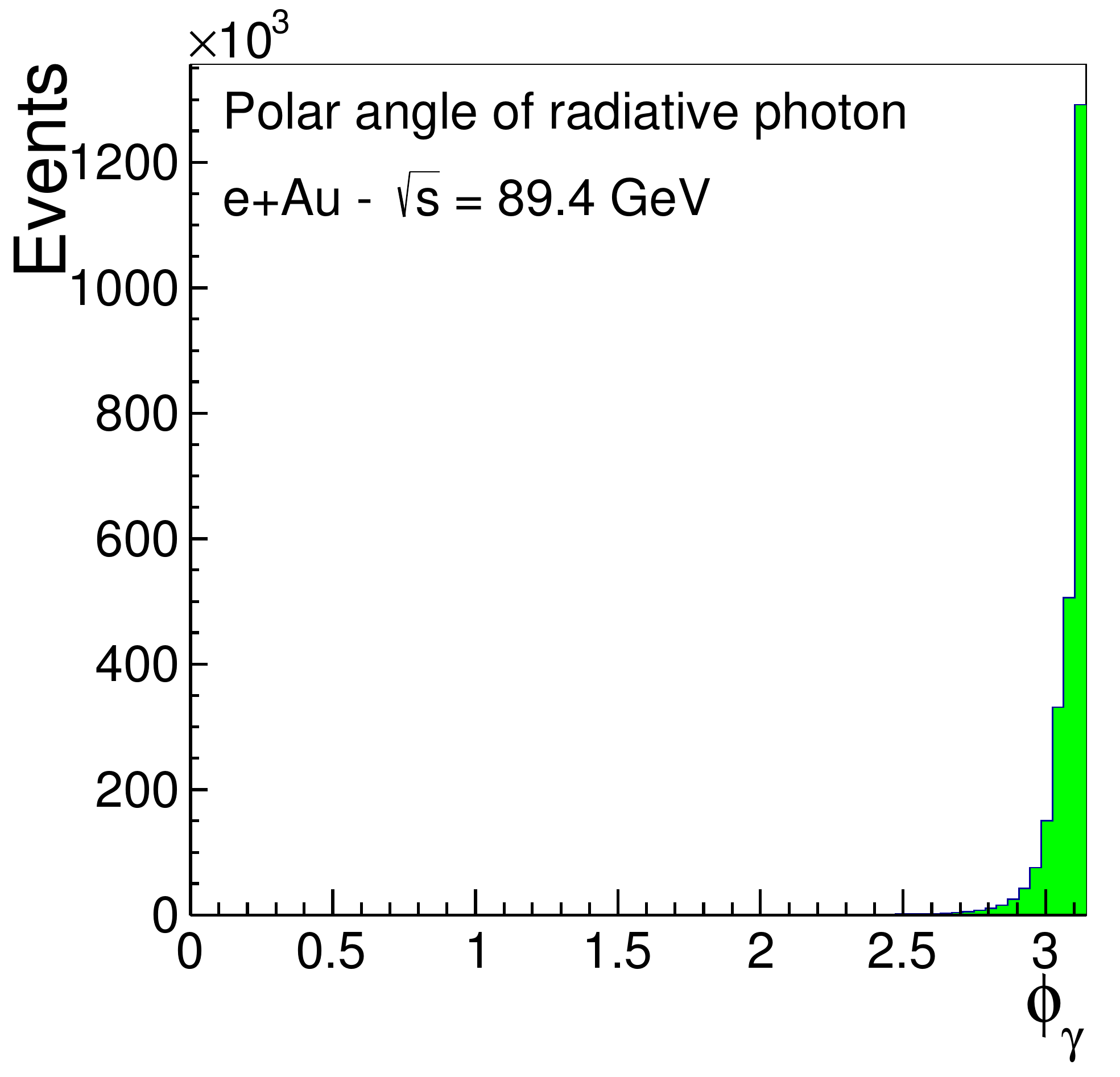}
   \caption{The energy ({\it left}) and polar angle ({\it right}) 
   distribution of radiative photons emitted in e+Au collision events.}
   \label{Fig:RadPhoton}
\end{figure}

The effect of the QED radiation off the incoming and outgoing lepton can be quantified by a correction 
factor
\begin{equation}
R_{\rm C}= \frac{\sigma_{\rm r}(\mathcal{O}(\alpha_{\rm em}))}{\sigma_{\rm r}(\rm born)} - 1,
\end{equation}
where $\sigma_{\rm r}(\rm born)$ and $\sigma_{\rm r}(\mathcal{O}(\alpha_{\rm em}))$ are the reduced cross 
section at born-level and including the first-order radiative corrections, respectively. 
To compute the above correction factors for $\sigma_{\rm r}$ and $\sigma_{\rm r}^{c\bar{c}}$ for the EIC kinematics, a sample of events were generated using the DJANGO simulator \cite{Schuler:1991yg}. The DJANGO Monte-Carlo generator was recently expanded to simulate $\ell$+$A$ collisions including $\mathcal{O}(\alpha_{\rm em})$ radiative effects. The simulations show that most of the radiative real photons have an energy much below 1~GeV, as shown in 
Figure~\ref{Fig:RadPhoton} ({\it left}). These radiative photons are typically emitted at very rear angles (in the electron going direction), see Figure~\ref{Fig:RadPhoton} ({\it right}), and are uniformly distributed in azimuthal angle. 

\begin{figure*}[htbp] 
   \centering
   \includegraphics[width=0.48\textwidth]{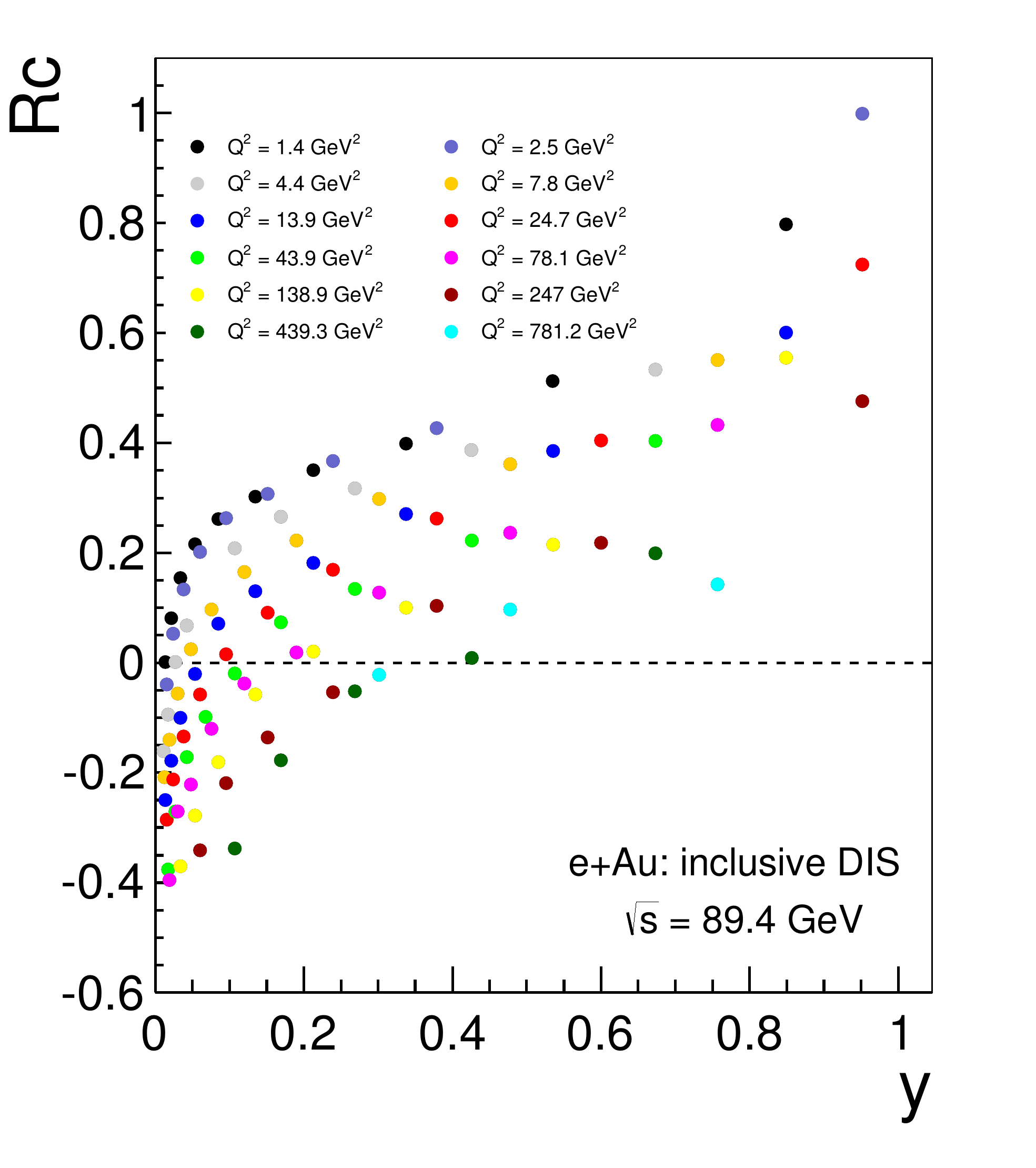} 
   \includegraphics[width=0.48\textwidth]{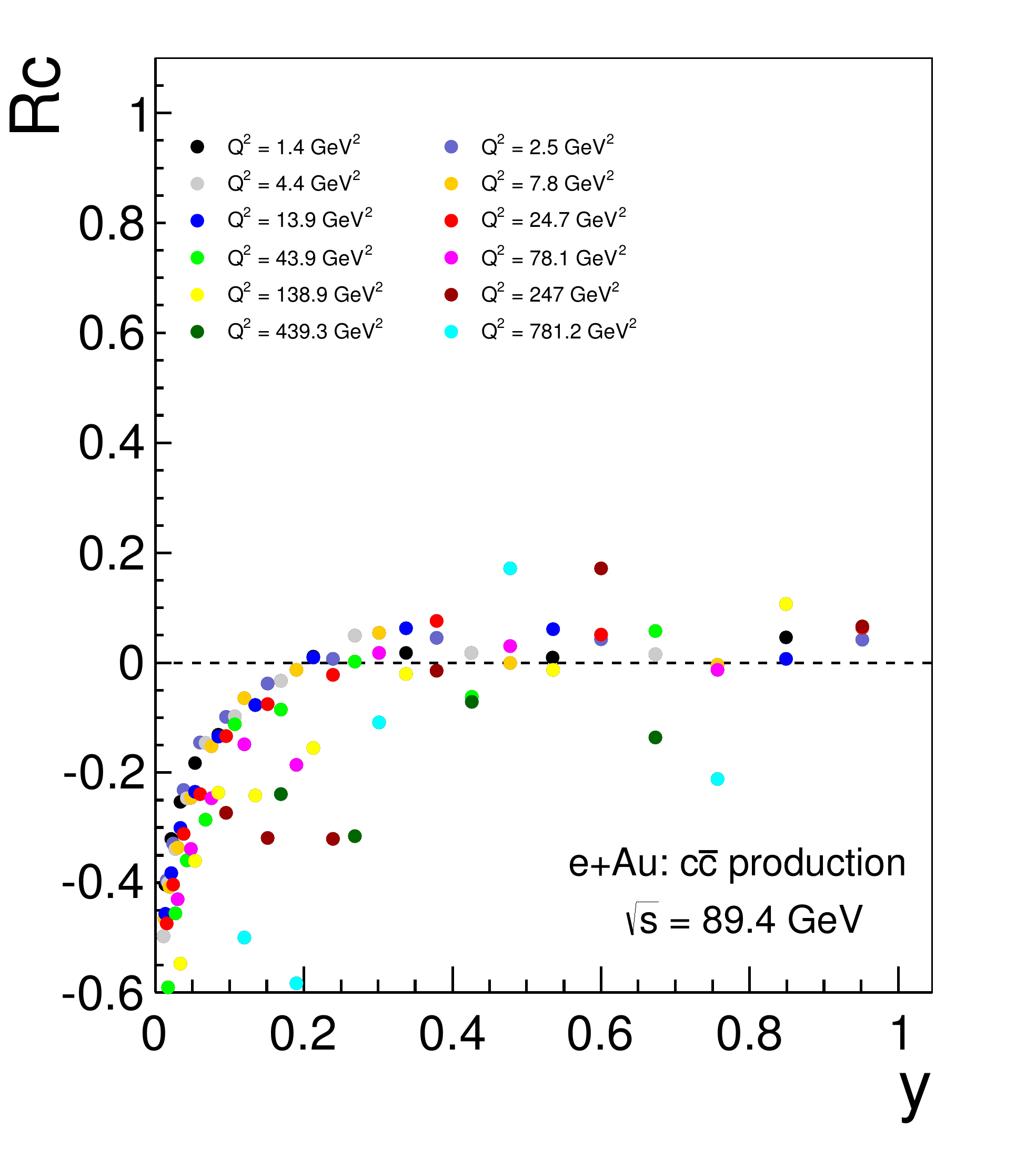}
   \caption{The radiative correction factor to the inclusive reduced cross section ({\it left}) and the 
reduced cross section for charm production  ({\it right}) as a function of the inelasticity, estimated for  
20~GeV electrons off 100~GeV Au-ions collision events, at different $Q^{2}$ values.}
   \label{Fig:Rc}
\end{figure*}

Figure~\ref{Fig:Rc} shows the radiative correction factor versus the inelasticity, $y$, due to QED 
radiation in $e$+Au collisions at $\sqrt{s}=89.4$~GeV for different $Q^{2}$ values, 
in the case of inclusive ({\it left plot}) and charm ({\it right plot}) reduced cross sections.
These values are compatible with earlier predictions~\cite{Boer:2011fh}. 
In the photon-nucleon center-of-mass frame, the maximum energy of the radiated photon, 
$E_{\gamma}^{\rm max}$, is given by
\begin{equation}
E_{\gamma}^{\rm max} / \sqrt{s}\approx \sqrt{y\left(1-\frac{Q^2}{sy}\right)} \label{eq:maxegamma}
\end{equation}
One can see that as the inelasticity $y$ grows, larger QED corrections are expected, on the other hand if 
 $Q^2$ grows, $E_{\gamma}^{\rm max}$ decreases and we can anticipate that the corrections get smaller. 
This behaviour can be verified from Figure~\ref{Fig:Rc} (left) in the case of
the inclusive cross section. Towards small $y$ and large $Q^2$ the phase space available for the photon 
emission becomes more and more restricted, and the correction factor falls strongly and becomes finally 
negative. This is a typical behavior if the phase space for photon emission becomes restricted 
and negative virtual corrections dominate (incomplete cancellation of infrared divergences).

The size of radiative corrections can be reduced utilizing information about the hadronic final state. 
Increasing the invariant mass of the hadronic final state leads to narrower phase space available for 
photon emission. 
This is the reason why the radiative correction factor for $\sigma_{\rm r}^{c\bar{c}}$ shown in 
Figure~\ref{Fig:Rc} is significantly reduced at high $y$ and high $Q^2$.
For $\sigma_{\rm r}$ a simple cut on the invariant mass of the hadronic final state $W_{\rm had}$ 
will reduce $R_C$.
A similar effect can be achieved cutting on $E - p_z$  ($p_z$: longitudinal momentum of hadronic final state 
particles) from the Jacquet-Blondel  
method \cite{Jacquet:1979jb,Bassler:1994uq}. The reduction of the radiative corrections will be 
considerable at largest $y$ and at small $x$, but probably not yet sufficient at larger values of $x$. 

\subsection{The longitudinal structure function}
\label{Sec:FL}

\begin{figure}[h] 
   \centering
   \includegraphics[width=0.48\textwidth]{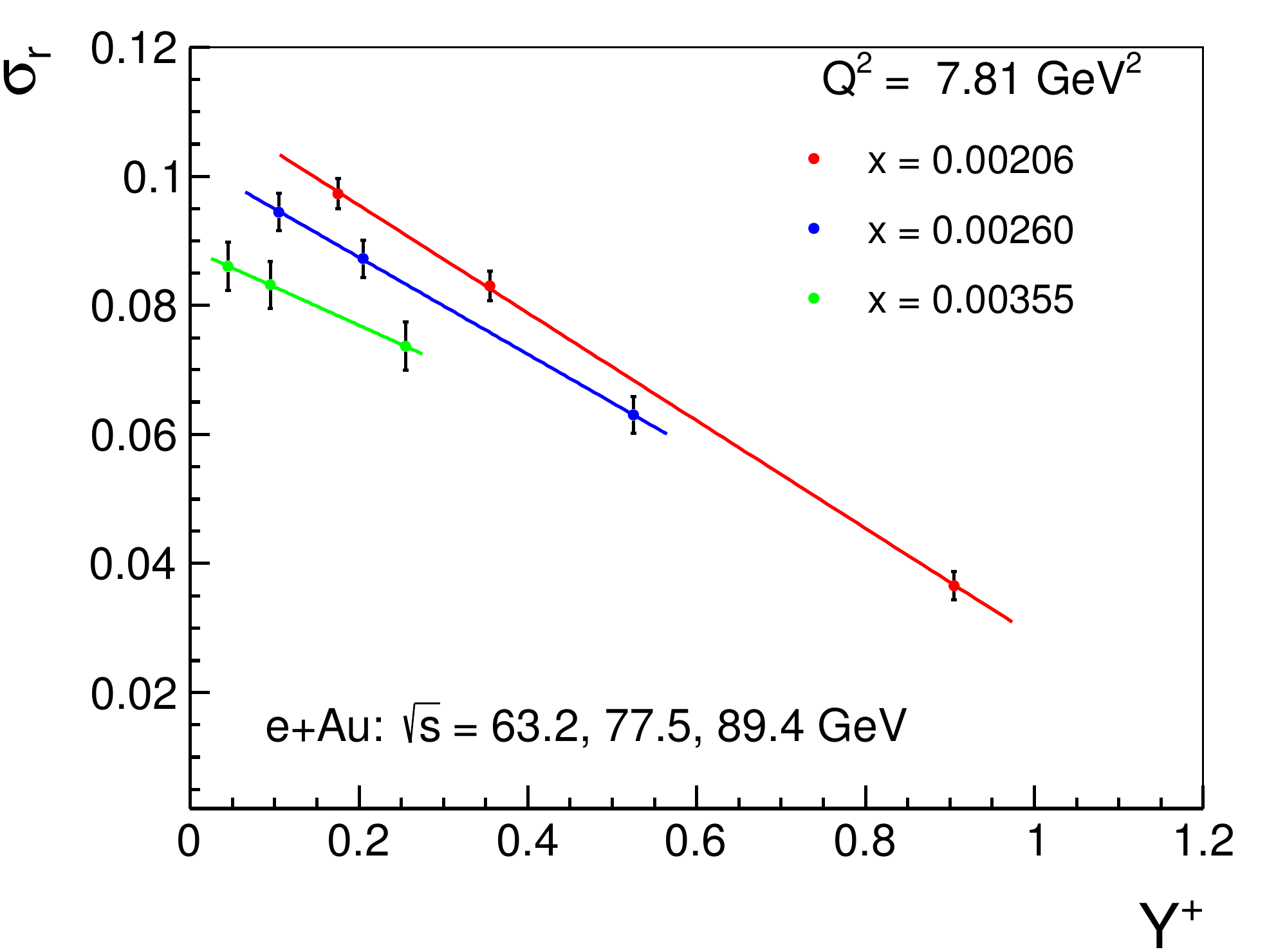}
   \caption{
   The reduced cross section for $e^-$+Au collisions at $\sqrt{s}= 63.2, 77.5~\text{and}~89.4$~GeV 
   versus $Y^+$ for $Q^2 = 7.81~\text{GeV}^2$ for three values of $x$. 
   Fitting the slope of each data set with fixed $x$ gives the negative of $F_L$.  }
   \label{Fig:Rosenbluth}
\end{figure}

As $F_{\rm L}$ is typically very small it is a demanding
quantity to study experimentally~\cite{Chekanov:2009na,Abramowicz:2014jak,Andreev:2013vha}. It is usually extracted 
through a Rosenbluth separation analysis. This requires measuring $\sigma_{\rm r}$ for at least three 
different c.o.m. energies and extracting $F_{\rm L}$ from a fit of $\sigma_{r}$  as function of 
$Y^{+} \equiv y^2/(1+(1-y)^2)$ for each bin in $x$. It is clear from Eq.~(\ref{Eq:StructFunc}), that the 
slope of this distribution represents $F_{\rm L}$.   
Therefore, having at hand data with enough range in c.o.m. energy to provide a good lever arm 
in $Y^{+}$ will be crucial for obtaining good-quality fits and extracting precise values of 
$F_{\rm L}$. 

To illustrate the extraction of $F_{\rm L}$ from $\sigma_{\rm r}$, Figure~\ref{Fig:Rosenbluth} shows the 
simulated $\sigma_{r}$ in $e^-$+Au collisions for 
$\sqrt{s}= 63.2, 77.5~\text{and}~89.4$~GeV at $Q^2=7.81$~GeV$^2$ as a function of $Y^{+}$ for three different 
$x$-values.  In order to ensure that the fit gives reasonable results, at least 3 points within a lever-arm in 
$Y^+$ larger than 0.1 are required. To compensate the collapsing lever-arm in $Y^+$ with increasing $x$, lower 
electron c.o.m. energies of $\sqrt{s}= 31.6, 38.7~\text{and}~44.7$~GeV are critical in order to reach higher-$x$.

\begin{figure*}[htbp!] 
   \centering
   \includegraphics[width=0.425\textwidth]{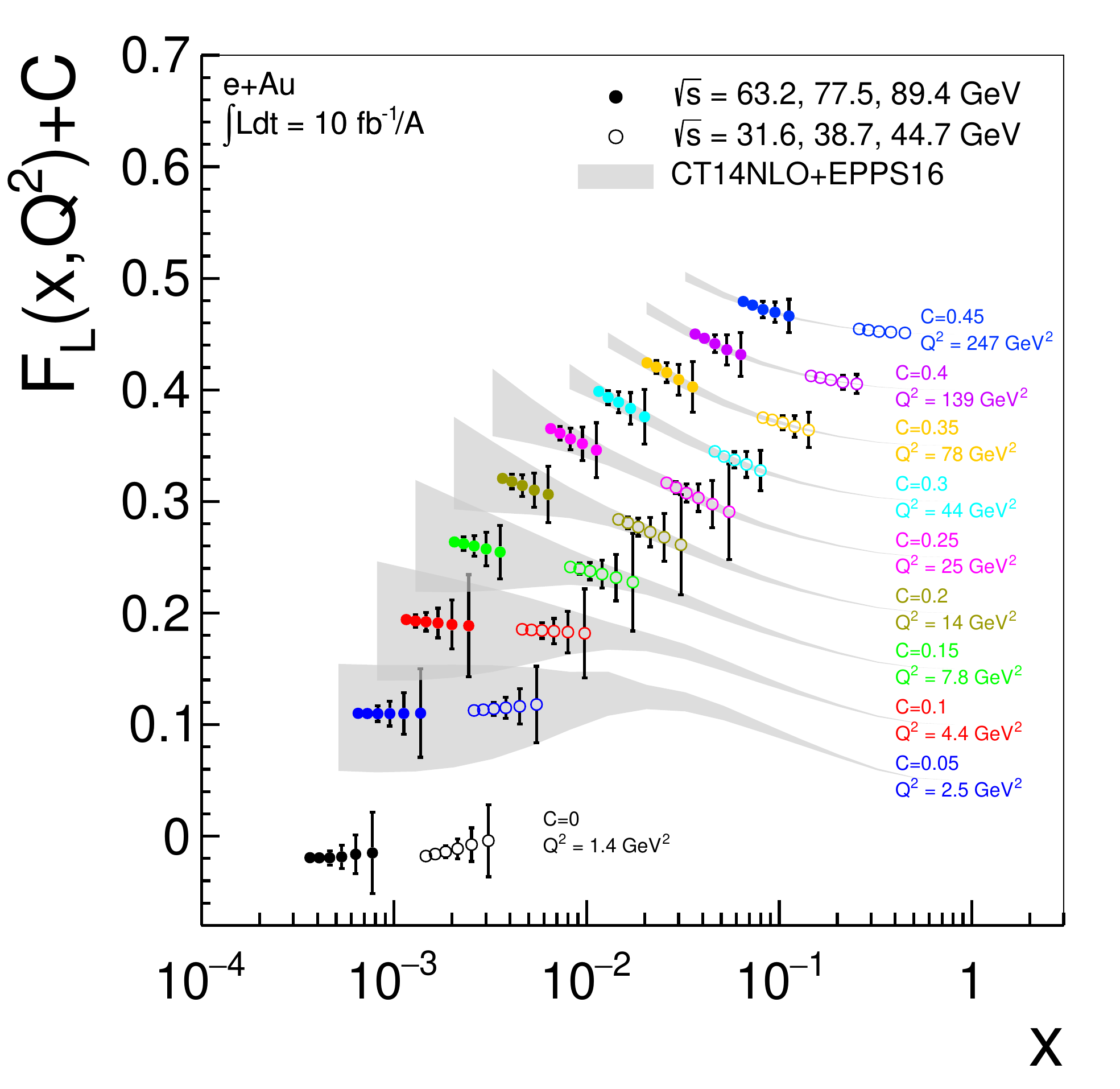}   
   \includegraphics[width=0.44\textwidth]{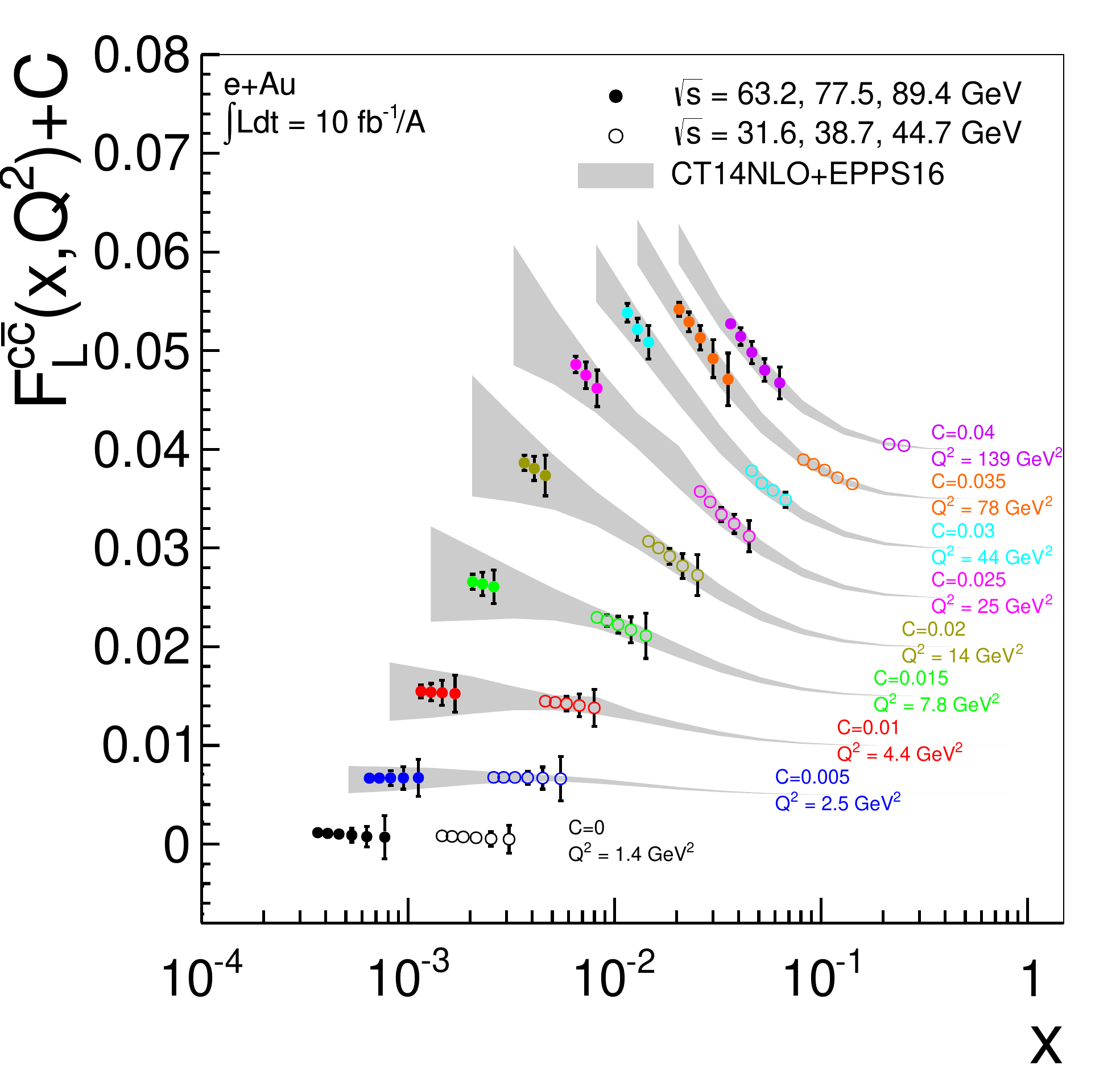}   
   \caption{Inclusive $F_L$ ({\em left}) and $F_L^{c\bar{c}}$ ({\em right}) as a function of $x$ for several 
    values of $Q^{2}$. The vertical bars represent statistical and systematic uncertainties added in quadrature.
    The grey bands represent the theoretical predictions based on EPPS16.}
   \label{Fig:FL}
\end{figure*}

As in the case of reduced cross sections, also for $F_{\rm L}$ we studied the potential of an EIC to measure both 
the inclusive and the charm structure functions in $e^-$+Au collisions. The collection of our results versus $x$ 
for a number of $Q^{2}$ values is shown in Figure~\ref{Fig:FL} for $F_{\rm L}$ ({\em left}) and 
$F_{\rm L}^{c\bar{c}}$ ({\em right}). The three different c.o.m. energies used in each extraction are also 
indicated on the plots. Measurements performed using a 5~GeV and a 20~GeV electron beam are indicated on the plots 
by open and solid circles respectively.
For clarity, the values are offset by adding a constant factor C. 
The NLO predictions using the CT14NLO free proton PDFs with the EPPS16 nuclear modifications are shown by the gray bands. One can see that, in comparison to the PDF error bands, with a combined collected luminosity of 10~fb$^{-1}$ at each electron beam-energy configuration, an EIC can perform a very precise measurement of the inclusive $F_{L}$ and $F_{L}^{c\bar{c}}$ in several $x, Q^{2}$ bins. With the highest c.o.m. energies, the longitudinal structure functions can be measured with a high precision down to $x \sim 7 \times 10^{-4}$ at low $Q^2$. At this low values of $x$, the predictions of saturation models for $F_{L}$ are already distinctively different from those of collinear factorization \cite{Marquet:2017bga}, which underscores the cruciality of such a mesurement.


It is also important to note that an EIC can achieve a comparable precision in measuring $F_{L}$ and $F_{L}^{c\bar{c}}$ for the 
proton, significantly improving the existing measurements from HERA~\cite{Chekanov:2009na,Andreev:2013vha}. 

%% file: MasterTexFiles/nPDFfit.tex
\section{Impact of an EIC on nuclear PDFs}
\label{sec:nPDFfit}

\subsection{Pseudodata for cross-section ratios}

To estimate the impact an EIC would have on nuclear PDFs, we have generated a sample of pseudodata $D_{i}$ for the ratios $\sigma_{\rm r}^{e^-+{\rm nucleus}}/\sigma_{\rm r}^{e^-+{\rm proton}}$. The pseudodata are based on a NLO calculation using the CT10NLO free proton PDFs and EPS09 for the nuclear modifications, denoted here by $T_{i}^{\rm EPS09}$. The values $T_i^{\rm EPS09}$ were distorted in the same way as in Ref.~\cite{d'Enterria:2015jna} by adding Gaussian noise according to the estimated percentual point-by-point uncorrelated ($\delta^{\rm uncorr.}_i$) and normalization uncertainties ($\delta^{\rm norm.}_i$) as
\begin{equation}
D_{i} = T_i^{\rm EPS09} \times \left[1 + \delta^{\rm uncorr.}_{i} r_{i} + \delta^{\rm norm.}_{i} r^{\rm norm.} \right], \label{eq:pseudodata}
\end{equation}
where $r_{i}$ and $r^{\rm norm.}$ are Gaussian random numbers with unit variance. The uncertainties from $e^-+A$ and $e^-+{\rm p}$, added in quadrature, 
are included in $\delta^{\rm uncorr.}_i$ and $\delta^{\rm norm.}_i$. The construction of pseudodata has been done independently for
each $\sqrt{s}$ and for two nuclei, Carbon ($^{12}$C) and Gold ($^{197}$Au). We have not accounted for any experimental correlations between the pseudodata for different $\sqrt{s}$ or different nuclei, as these are difficult to estimate at this stage. The luminosity uncertainty of 1.4\% is assumed to be uncorrelated for each data set with different $\sqrt{s}$ and nucleus, and it is treated in the global $\chi^2$ minimization as in the EPPS16 analysis \cite{Eskola:2016oht}. For clarity, in Table~\ref{Tab:Uncert} we record the assumed values for systematic uncertainties discussed already erlier in Sec.~\ref{sec:MonteCarlo}. 

\begin{table}[htbp!]
\caption{The systematic uncertainties of inclusive and charm-tagged cross-section measurements. The values are in percents.}
\label{Tab:Uncert}
\begin{center}
\begin{tabular}{|c|c|c|c|}
\hline
Sources of Uncertainty & Value in $\sigma_{r}$ (\%) & Value in $\sigma_{r}^{c\bar{c}}$ (\%) \\
\hline
Luminosity            & 1.4 & 1.4 \\
Electron id. and eff. & 1.6 & 1.6 \\
RICH and $dE/dx$ PID  & 0 & 3     \\
Vertex finding        & 0 & 1     \\
\hline
\end{tabular}
\end{center}
\end{table}

\subsection{nPDF analysis}

\begin{figure}[htb!]
\center
\includegraphics[width=0.42\textwidth]{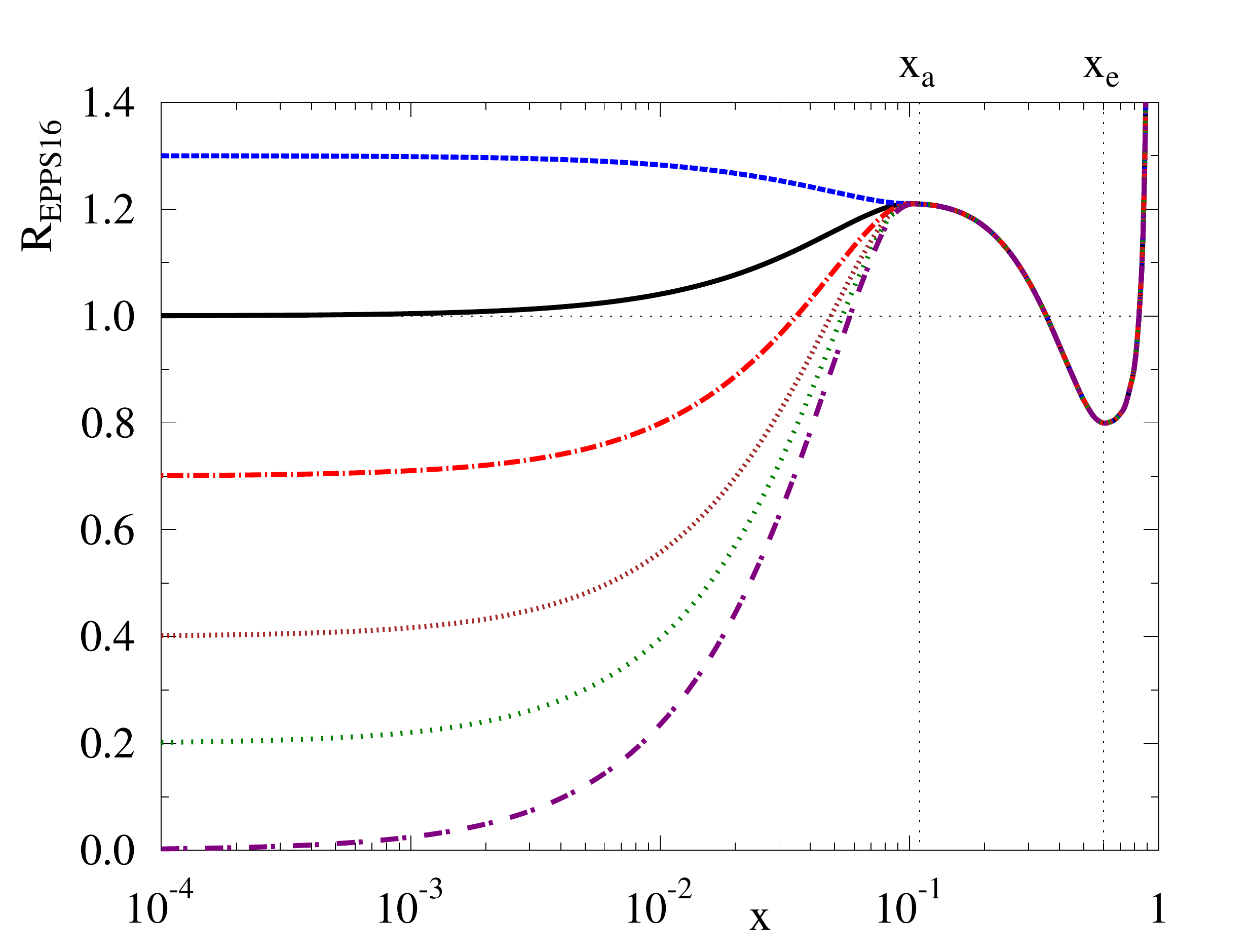}
\includegraphics[width=0.42\textwidth]{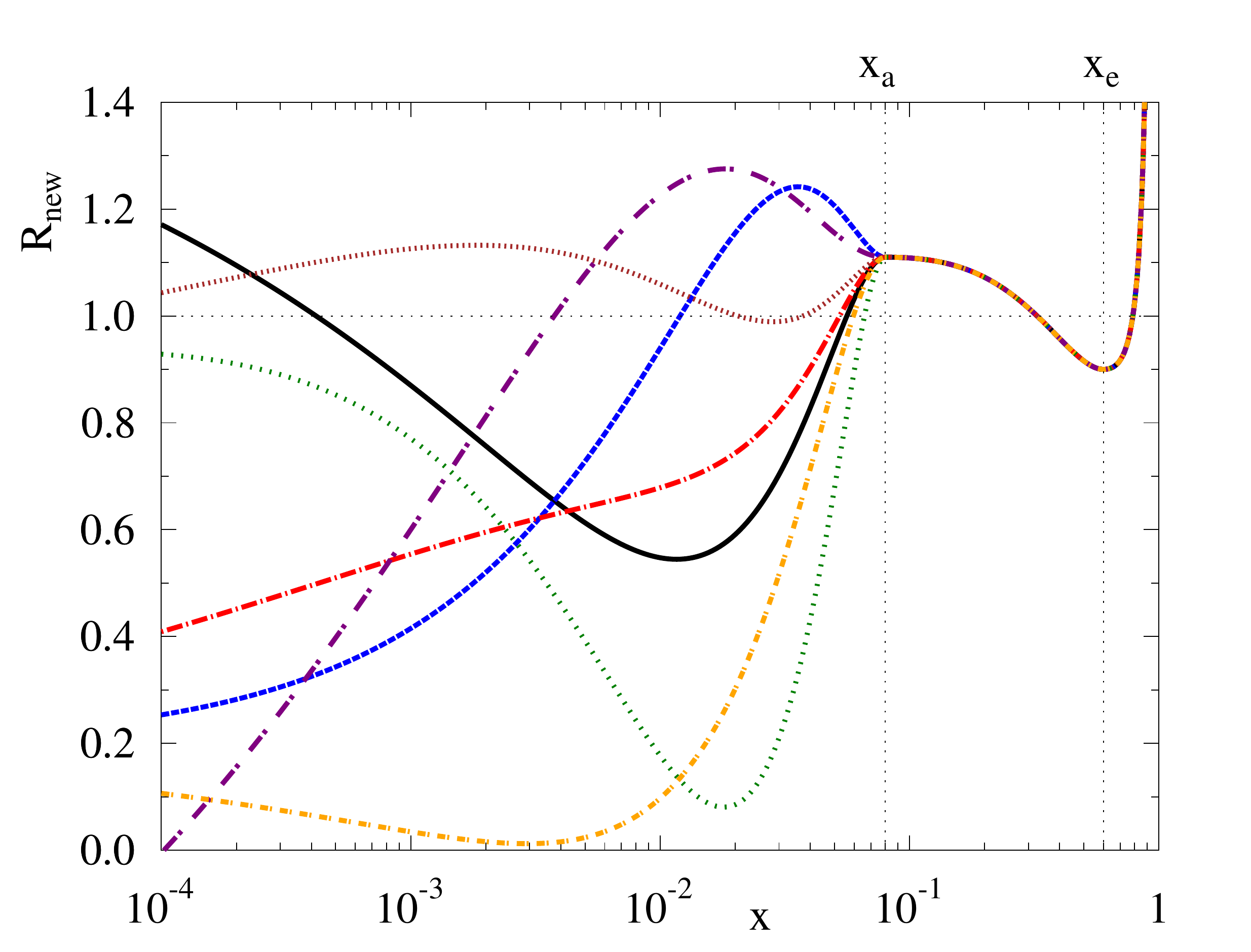}
\caption{Illustration of the rigidity/flexibility of the small-$x$ fit functions used in EPPS16 analysis 
(\emph{upper}) and in the present work (\emph{lower}).}
\label{fig:Fitfunction}
\end{figure}

As an EIC would extend the current kinematic reach of $e^-$+$A$ measurements to smaller values of $x$, it is 
clear that such new information would have an impact on the global extractions of nPDFs. One way to quantitatively 
address the improvement that an EIC would entail is to take advantage of PDF re-weighting techniques 
\cite{Paukkunen:2014zia}. However, once the new measurements probe the PDFs in a previously unconstrained kinematic 
range, care has to be taken that the results are not overly affected by parametrization bias. Here, our starting 
point is the recent global analysis of nPDFs, EPPS16 \cite{Eskola:2016oht}. There, the nuclear modification of the
proton PDF is defined as 
\begin{equation}
R_i(x,Q^2) \equiv \frac{f_i^{{\rm proton}/A}(x,Q^2)}{f_i^{{\rm proton}}(x,Q^2)},
\end{equation}
where $f_i^{{\rm proton}/A}(x,Q^2)$ denotes the bound-proton PDF for flavor $i$ and $f_i^{{\rm proton}}(x,Q^2)$ is the corresponding free-proton PDF. 
The adopted $x$ dependence was 
\begin{equation}
R_{\rm EPPS16}(x) = 
\left\{
\begin{array}{lc}
a_0 + a_1(x-x_a)^2 & x \leq x_a \\
b_0 + b_1x^\alpha + b_2x^{2\alpha} + b_3x^{3\alpha} & x_a \leq x \leq x_e \\
c_0 + \left(c_1-c_2x \right) \left(1-x\right)^{-\beta} & x_e \leq x \leq 1.
\end{array}
\right.  \label{EPPS16R}
\end{equation}
In the equations above, $x_a$ and $x_e$ are the values of $x$ corresponding to the assumed antishadowing maximum 
and EMC minimum, respectively (see Figure~\ref{fig:Fitfunction}). The rest of the parameters were adjustable but 
constrained such that the piecewisely defined parametrization is smooth over all $x$.
The $A$ dependence of the fit functions was encoded with a power-law-like parametrization at $x=x_a$, $x=x_e$, 
and in the case of sea quarks also in the limit $x\rightarrow 0$, see Ref~\cite{Eskola:2016oht} for further details. 
Figure~\ref{fig:Fitfunction} (\emph{left}) shows some examples of how the function in Eq.~\eqref{EPPS16R} behaves at 
small $x$ when freezing the parameters that control the region $x>x_a$.
The stiffness of $R_{\rm EPPS16}(x)$ is obvious: only a monotonic decrease or increase towards $x\rightarrow 0$ is 
possible. Exactly the same limitation would apply also if we were to perform a PDF-reweighting study. 
Here, our goal is to partly release this assumption to obtain a less-biased estimate of the projected data 
constraints. In practice, we have replaced the EPPS16 small-$x$ fit function in Eq.~\eqref{EPPS16R} by a more 
flexible form.
\begin{equation}
R_{\rm new}(x \leq x_a) = a_0 + (x-x_a)^2 
    \left[ a_1 +
    \sum_{k=1}^2 a_{k+2}x^{k/4}
    \right]. \label{eq:newf}
\end{equation}
Some examples of how this function can behave are shown in Figure~\ref{fig:Fitfunction} (\emph{right}).
Ideally, the same functional form should be applied to all partonic species, but in the present work we only use it for the gluons. They arguably play a special role being particularly prone to non-linear effects at low $Q^2$ and also in controling the small-$x$ behaviour of sea quarks at higher $Q^2$ through $g\rightarrow q{\overline q}$ splitting. In fact, an extension to all parton flavors would require a complete change in the analysis methodology which is beyond the scope of this work. 
This is because the Hessian method \cite{Pumplin:2001ct} that was used in the EPPS16 analysis (and is used here, 
too) to quantify the PDF uncertainties becomes unstable in the presence of large uncertainties and complex 
correlations among the fit parameters (within a single flavor and across various flavors). To overcome this 
limitation, Monte-Carlo techniques \cite{Giele:1998gw,Giele:2001mr,Watt:2012tq} should be used instead. 
This is left as a future work.

After adopting the more flexible functional form for the gluons also the baseline, against which the effect of 
an EIC should be contrasted, will be different from EPPS16. Thus, we have also performed a global nPDF fit, 
which is otherwise equal to the EPPS16 analysis, with the exception of the more flexible functional form of 
Eq.~(\ref{eq:newf}) to parametrize the small-$x$ gluon nuclear modifications. As in EPPS16, the nPDF uncertainties 
are determined via the Hessian method \cite{Pumplin:2001ct}, but in the present work the Hessian matrix is 
computed using the linearized prescription \cite{Martin:2009iq}. A fixed tolerance $\Delta \chi^2=50$, which 
corresponds approximately to the 90\% confidence-level of EPPS16, was employed. We have not repeated the 
determination of the 90\% confidence-level $\Delta \chi^2$ for all fits separately, though adding 
new data sets, especially with a large number of data points, has been observed to influence $\Delta \chi^2$ when 
it is computed on the basis of dynamical tolerance criterion. For example, in the EPPS16 analysis adding 
$\sim 900$ data points led to a 15-unit increase. However the uncertainty bands scale as $\sqrt{\Delta \chi^2}$, 
and no dramatic differences are expected from corrections on $\Delta \chi^2$.

\begin{figure*}[htbp!]
\center
\includegraphics[width=0.75\textwidth]{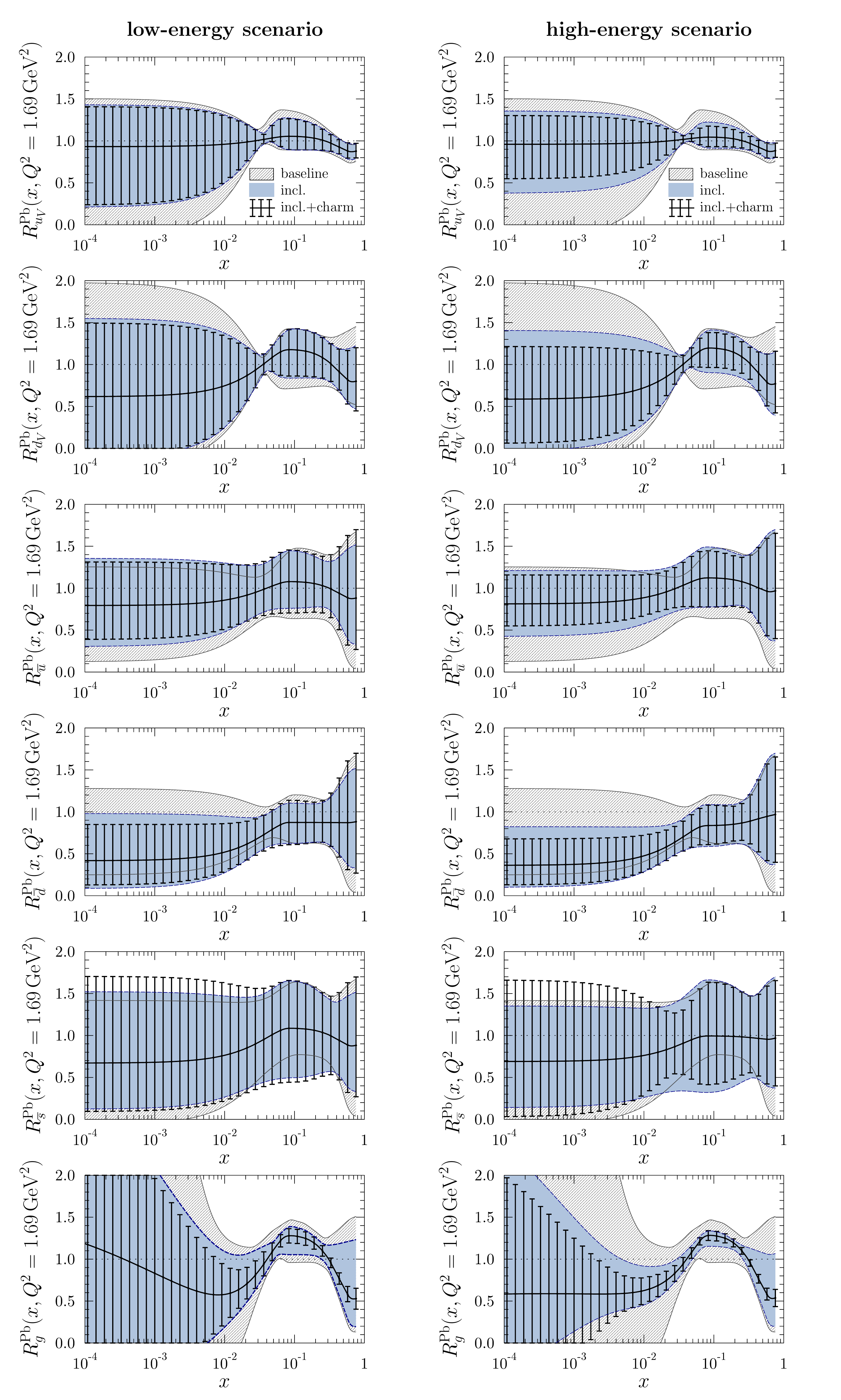}
\caption{Results for the nuclear modifications of Pb at $Q^2=1.69\,{\rm GeV}^2$. The hatched bands correspond to the baseline fit, the blue bands are the results from fits with no charm data included, and the black error bands denote the full analysis with inclusive and charm data.}
\label{fig:fit1}
\end{figure*}

\begin{figure*}[htbp!]
\center
\includegraphics[width=0.75\textwidth]{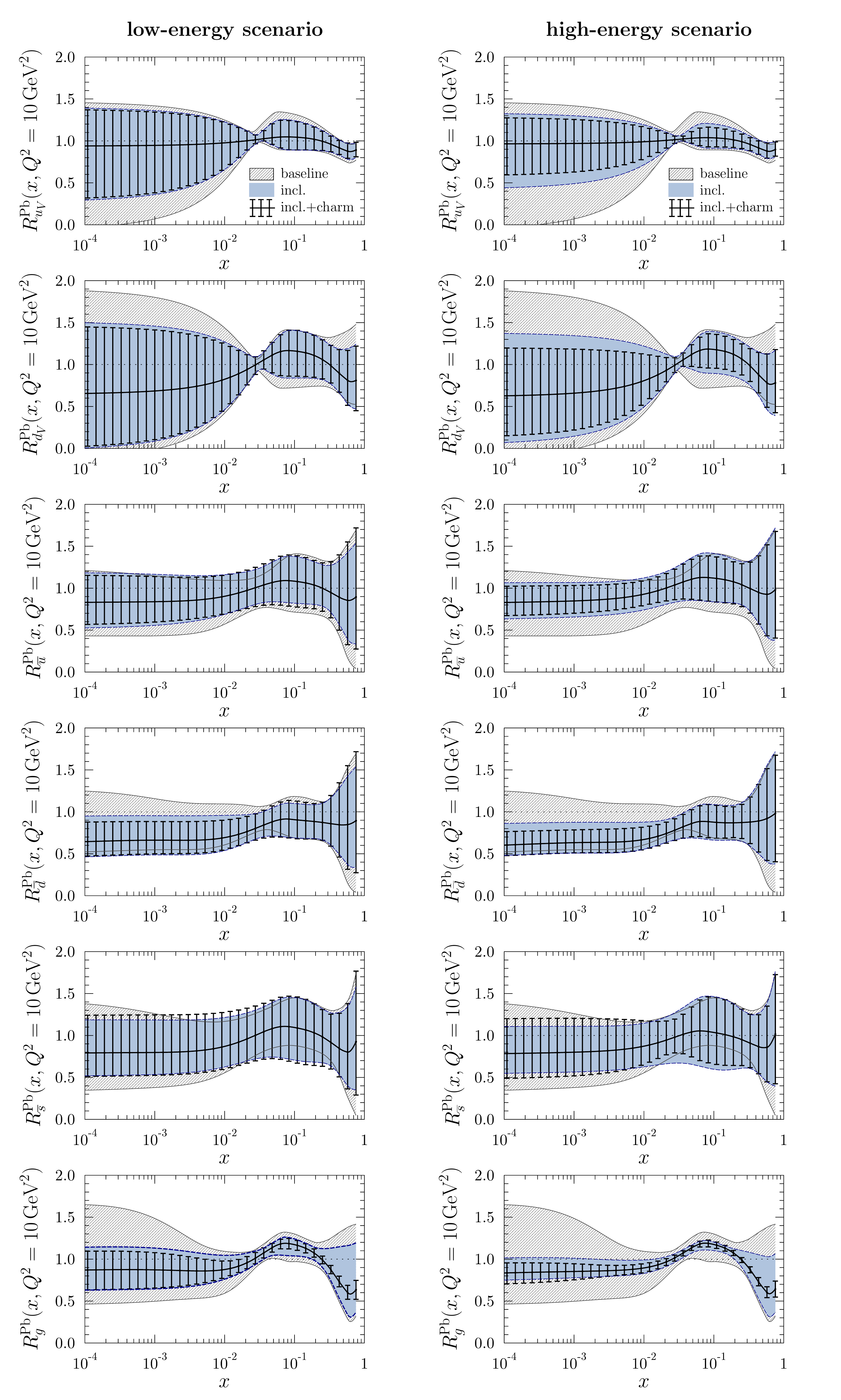}
\caption{As Figure~\ref{fig:fit1} but at $Q^2=10 \, {\rm GeV}^2$.}
\label{fig:fit3}
\end{figure*}

\begin{figure*}[htbp!]
\center
\includegraphics[width=0.75\textwidth]{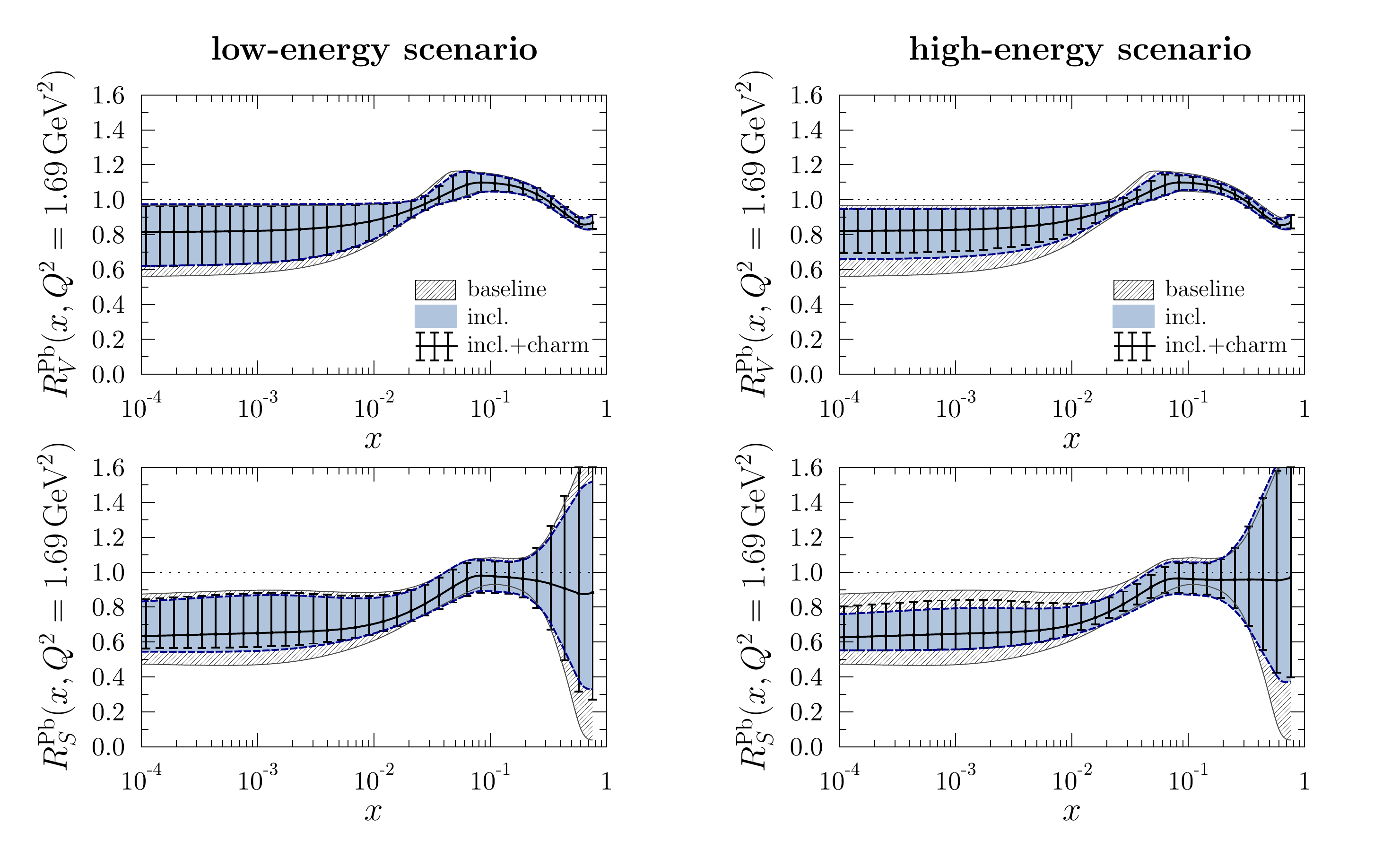}
\includegraphics[width=0.75\textwidth]{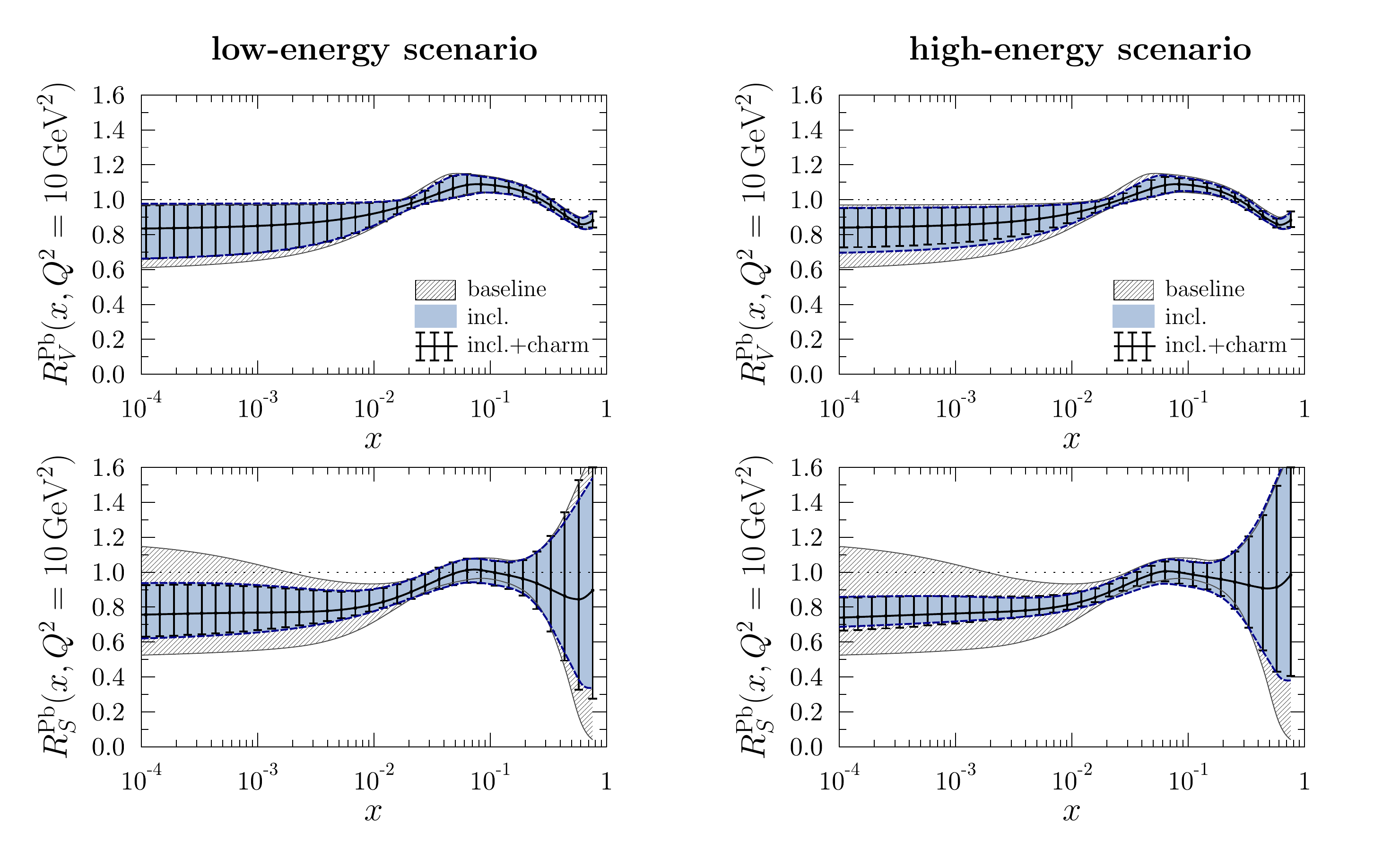}
\caption{As Figure~\ref{fig:fit1}, but for average valence (upper panels) and average light sea quarks (lower panels). The upper set of four panels corresponds to $Q^2=1.69\,{\rm GeV}^2$ and the lower sset of panels to $Q^2=10\,{\rm GeV}^2$.}
\label{fig:fit2}
\end{figure*}

We have studied the impact of various combinations of the EIC pseudodata: grouping different $\sqrt{s}$, using only 
the inclusive pseudodata, and incorporating the charm-tagged observables in addition. For the inclusive case the 
following energy configurations were used: \\

\begin{tabular}{lll}
$(E_{\rm e}/{\rm GeV},E_{\rm p,C,Au}/{\rm GeV})$&=&(5,50), (5,75), (5,100),\\
                                                & &(20,50), (20,75), (20,100),\\
\end{tabular}\\

\noindent and the charm pseudodata correspond to the setups\\

$(E_{\rm e}/{\rm GeV},E_{\rm p,C,Au}/{\rm GeV})$=(5,100), (20,100).\\

\noindent While also other combinations are possible, this collection already gives a good idea of the impact. 
In the following the data with $5 \, {\rm GeV}$ electron beam are referred to as ``low-energy scenario'' and the 
$20 \, {\rm GeV}$ electron beam data as ``high-energy scenario''. 

Figure \ref{fig:fit1} shows the resulting nuclear modifications for all partonic flavors at $Q^2=1.69\,{\rm GeV}^2$ (the parametrization scale), and Figure \ref{fig:fit3} at $Q^2=10\,{\rm GeV}^2$ (relevant e.g. for J/$\Psi$ production). The results are given for $^{208}$Pb nucleus which is the most relevant one for the current LHC heavy-ion program. The hatched bands represent the uncertainties from the baseline fit, the blue bands correspond to fits with inclusive data only, and the black bars to the analyses including the charm cross sections. The results are shown for both the low- and high-energy versions of an EIC. In the case of up and down quarks, the trends are quite clear --- the more data are used the narrower the uncertainty bands get,
up to a factor of two reduction at small $x$. 
In the case of valence quarks, this is a reflection of the fact that they get better constrained at $x \sim 0.1$, which also leads to smaller 
uncertainties  at small $x$ due to the valence-quark sum rules and form of the fit function. 
Neither the inclusive nor the charm cross sections are sensitive to the (anti-)strange quarks. As a result, there are no significant differences 
in the obtained nuclear modifications. To constrain (anti-)strange quarks at an EIC, measurements of charm production in charged-current reactions 
(mediated by $W^-$) or, perhaps, semi-inclusive kaon production should be considered.
For the gluons, the 
widths of the uncertainty bands evolve as expected: in the baseline fit the uncertainties are
rather significant at all values of $x$ and adding the inclusive EIC pseudodata brings the uncertainties down especially at small and mid $x$. 
Finally, when the charm-tagged pseudodata are incorporated into the analysis, the mid- and large-$x$ gluons become very well determined. 
It is stressed that the nucleons have been assumed to carry zero intrinsic charm at the charm mass treshold $Q^2=1.69\,{\rm GeV}^2$ --- all the 
charm quarks are generated perturbatively. Allowing a non-zero charm-quark content at the starting scale would 
presumably reduce the impact on the gluon uncertainty. 
However, the experimental evidence suggests that, in practice, the charm content of the nucleons at its mass threshold should be very small \cite{Ball:2017nwa}. In all cases, the nPDF extraction including the high-energy data has further reduced the uncertainties compared to the fit with 
low-energy data only. On one hand, the kinematic reach in $(x,Q^2)$ is better with high energies, see Figure~\ref{Fig:Kine-space}. On the other hand, 
the high-energy data has approximately twice as much data points, which also explains part of the improvements.

At first sight, it may appear puzzling that the uncertainties of up- and down-sea quark distributions at small-$x$ remain rather sizable in comparison to the typical $\sim$2\% uncertainty of the small-$x$ inclusive cross section data. The reason is the significant anticorrelation between the two quark flavors, which leads to cancellations in the 
cross sections: a larger $\overline{u}$-density is compensated by smaller $\overline{d}$-density and vice versa.
The situation can be illustrated
by examining the total nuclear $\overline{u}$-quark distribution $\overline{u}^A$ (for clarity the $x$ and $Q^2$ arguments are suppressed),
\begin{equation}
\overline{u}^A = \frac{Z}{A} R_{\overline{u}} \overline{u}^{\rm proton} + \frac{A-Z}{A} R_{\overline{d}} \overline{d}^{\rm proton} \, ,
\end{equation}
where $A$ is the nuclear mass number, and $Z$ the number of protons. We can decompose $\overline{u}^A$ in terms of the average modification $R_{\overline{u}+\overline{d}}$,
\begin{equation}
R_{\overline{u}+\overline{d}} \equiv ({R_{\overline{u}} \overline{u}^{\rm proton} + R_{\overline{d}} \overline{d}^{\rm proton}})/({\overline{u}^{\rm proton} + \overline{d}^{\rm proton}})
\end{equation}
and the difference $\delta R_{\overline{u}-\overline{d}}$,
\begin{equation}
 \delta R_{\overline{u}-\overline{d}} \equiv R_{\overline{u}} - R_{\overline{d}},
\end{equation}
 as
\begin{equation}
\begin{aligned}
\overline{u}^A & = & R_{\overline{u}+\overline{d}} \left( \frac{Z}{A} \overline{u}^{\rm proton} + \frac{A-Z}{A} \overline{d}^{\rm proton}\right) + \\
        & & \delta R_{\overline{u}-\overline{d}} \left( \frac{2Z}{A} - 1\right) \frac{\overline{u}^{\rm proton}}{1+ \overline{u}^{\rm proton}/\overline{d}^{\rm proton}} \,.
        \label{eq:totaluva}
\end{aligned}
\end{equation}
For an isoscalar nucleus (like $^{12}$C), the last term in Eq.~\eqref{eq:totaluva} is zero and thus the cross sections are not sensitive to the flavour separation. 
For non-isoscalar nuclei (like $^{197}$Au) the last term in Eq.~\eqref{eq:totaluva} is non-zero, but merely a correction to the leading term proportional to $R_{\overline{u}+\overline{d}}$. Indeed, at small $x$, $\overline{u}^{\rm proton} \approx \overline{d}^{\rm proton}$, and the term proportional to $\delta R_{\overline{u}-\overline{d}}$ is suppressed by a factor 
of $(Z/A-1/2) \approx -0.1$ for $^{197}$Au in comparison to the $R_{\overline{u}+\overline{d}}$ term. As a consequence, the sensitivity to flavor decomposition is always reduced in inclusive cross sections. A similar reasoning 
applies also in the case of valence quarks and explains the poor flavor decomposition 
at large $x$ despite the addition of high-precision EIC pseudodata.
To gain a better sensitivity to the flavor decomposition, wisely chosen differences of 
cross sections or structure functions in neutral- and charged-current reactions could lead to a partial cancellation of the $R_{\overline{u}+\overline{d}}$ terms, 
thereby increasing the importance of $\delta R_{\overline{u}-\overline{d}}$ terms (or equivalent for valence quarks). 

\begin{figure*}[htbp!]
\center
\includegraphics[width=1.0\textwidth]{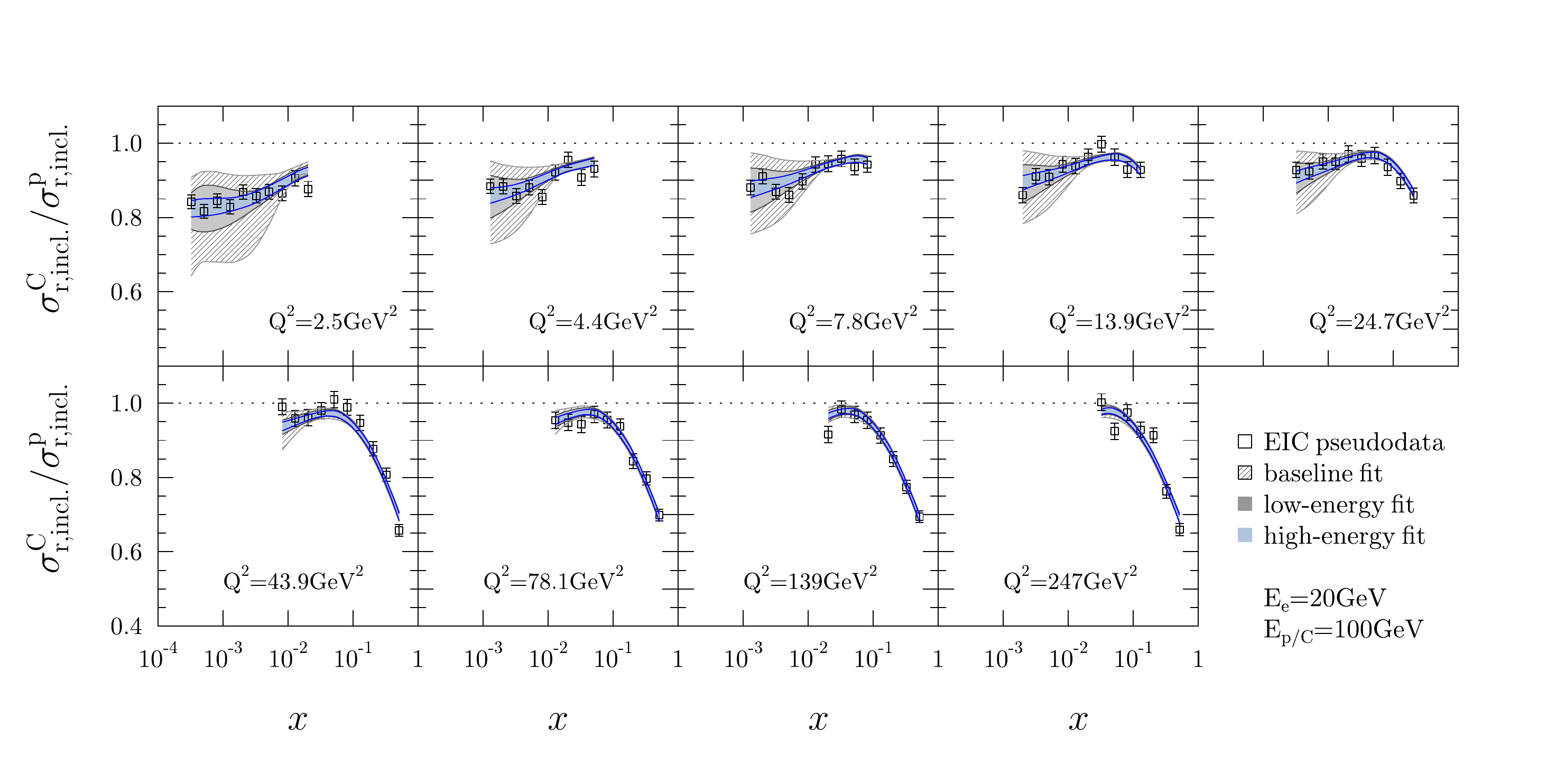}\vspace{-1.4cm}
\includegraphics[width=1.0\textwidth]{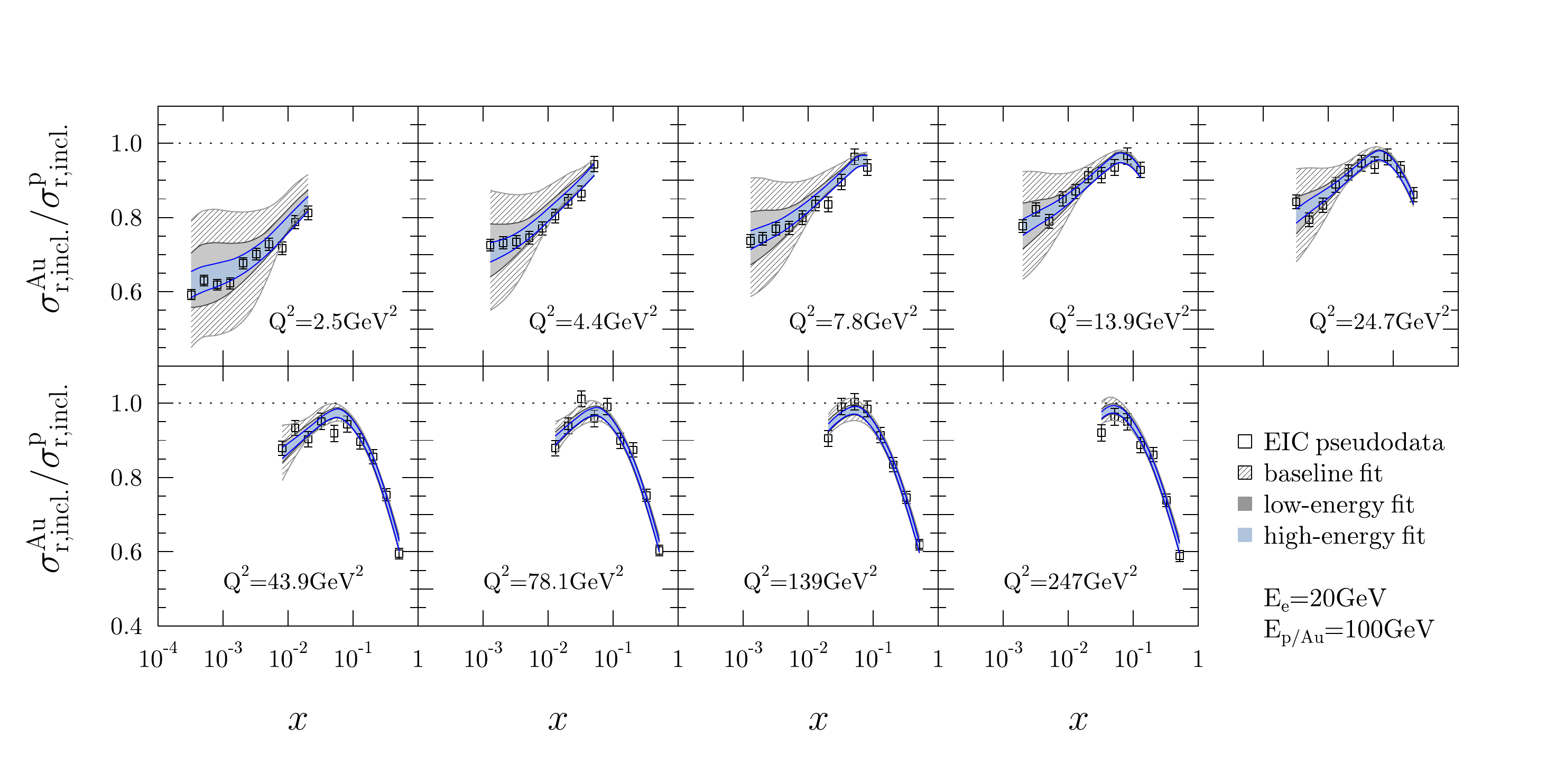}
\caption{The inclusive EIC pseudodata (in $E_{\rm e}=20\,{\rm GeV}$, $E_{\rm p,C,Au}=20\,{\rm GeV}$ setup) for Carbon (upper panels) and Gold (lower panels) compared with the baseline fit (hatched bands), the fit with inclusive low-energy data only (gray bands), and the fit with the inclusive low- and high-energy data (blue bands). The assumed overall 1.4\% data normalization uncertainty is not shown.}
\label{fig:datacomp1}
\end{figure*}

\begin{figure*}[htbp!]
\center
\includegraphics[width=1.0\textwidth]{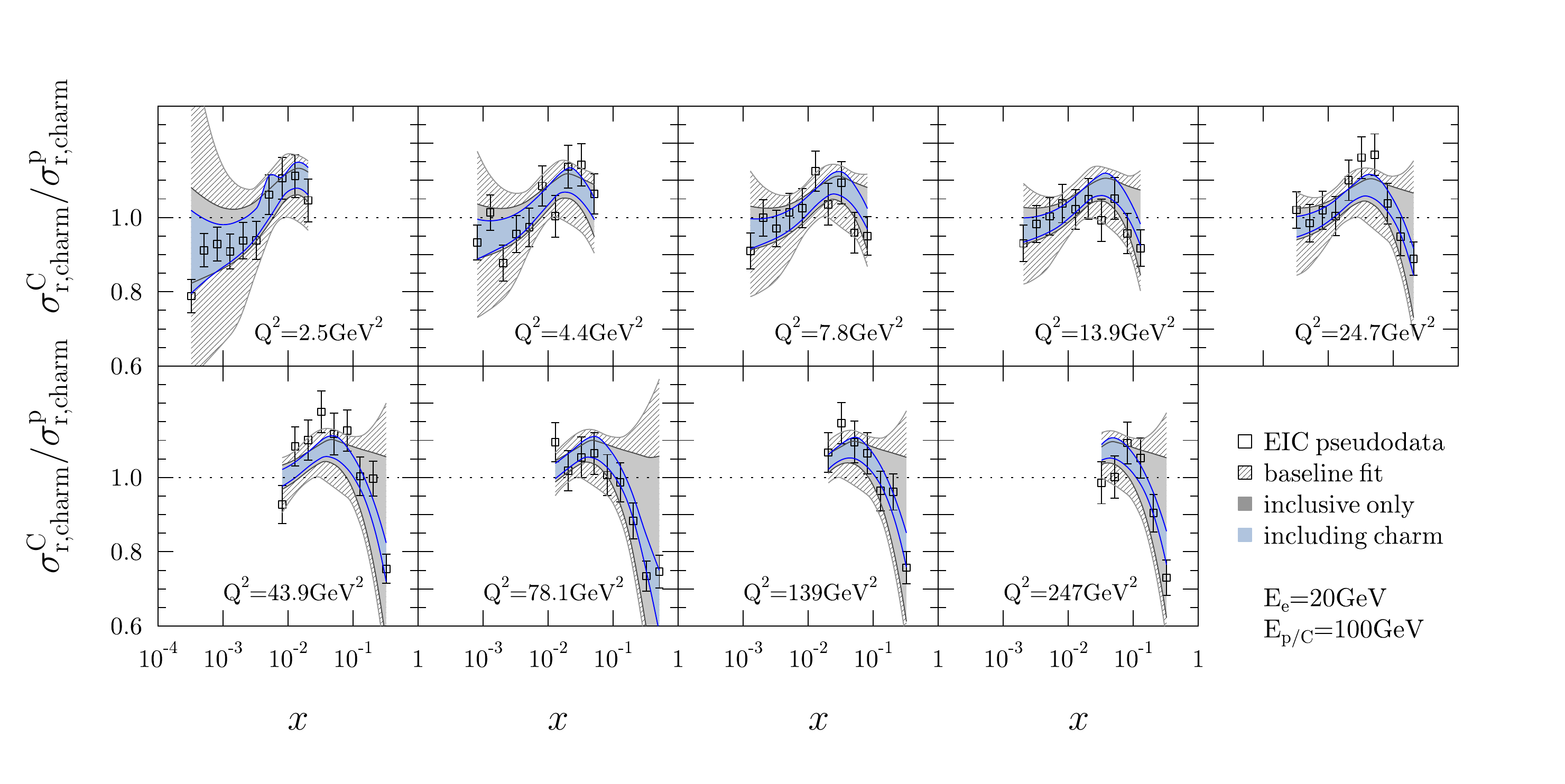}\vspace{-1.4cm}
\includegraphics[width=1.0\textwidth]{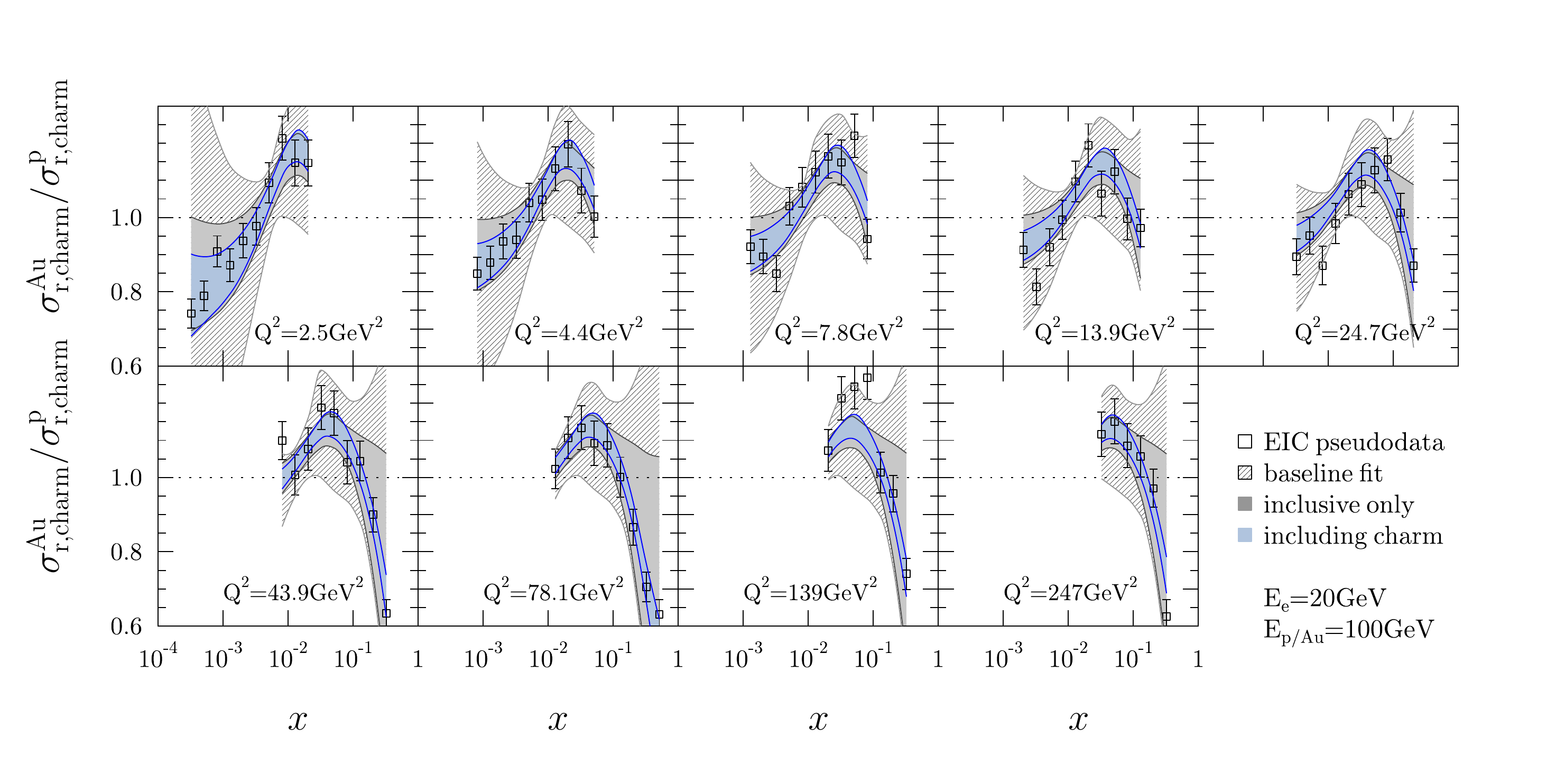}
\caption{The charm-tagged EIC pseudodata (in $E_{\rm e}=20\,{\rm GeV}$, $E_{\rm p,C,Au}=20\,{\rm GeV}$ setup) for Carbon (upper panels) and Gold (lower panels) compared with the baseline fit (hatched bands), the fit with inclusive EIC data included (gray bands), and charm data included (blue bands). The overall 1.4\% normalization uncertainty is not shown.}
\label{fig:datacomp2}
\end{figure*}

\begin{figure*}[htbp!]
\centering
\includegraphics[width=0.32\linewidth]{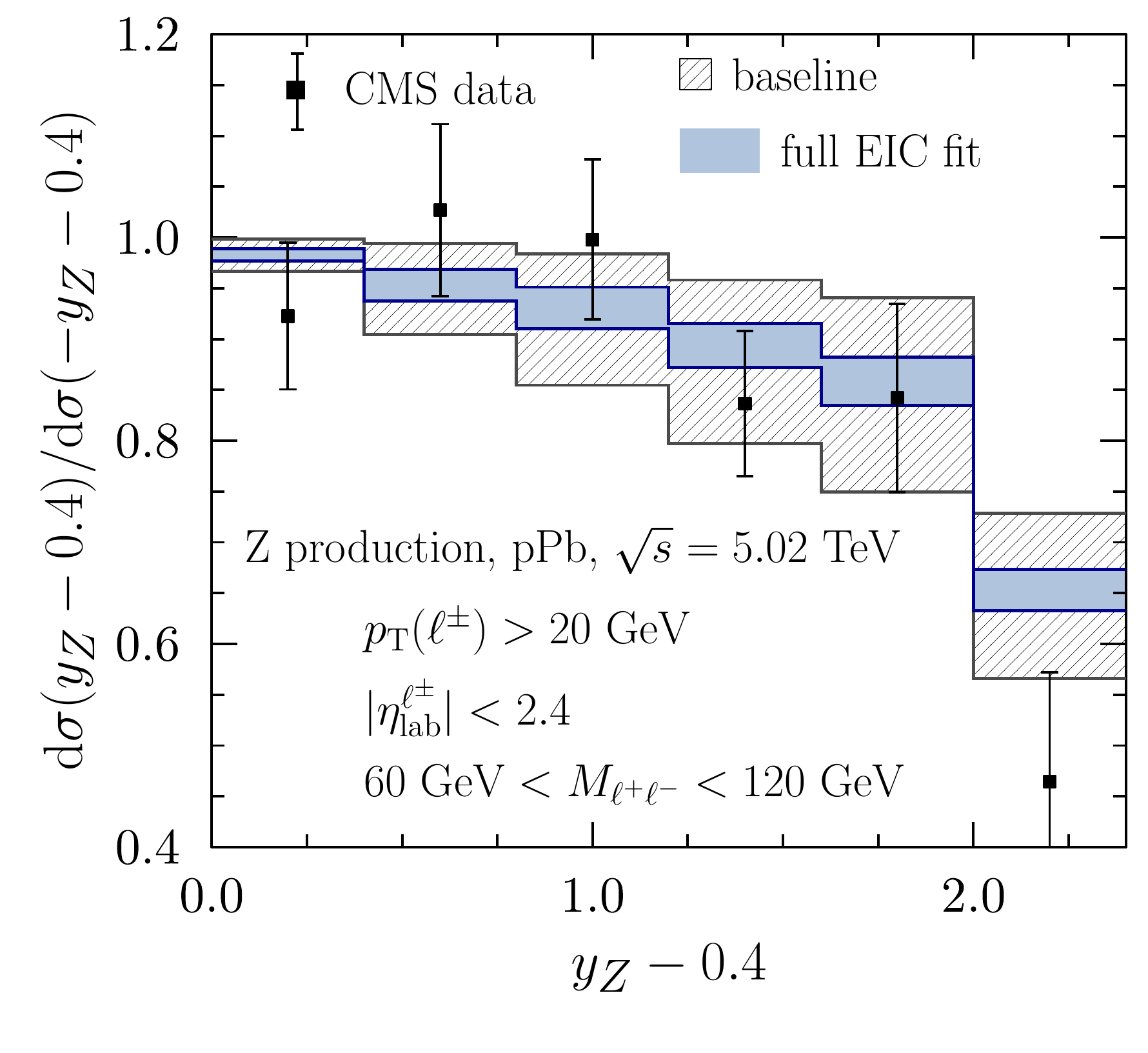}
\includegraphics[width=0.32\linewidth]{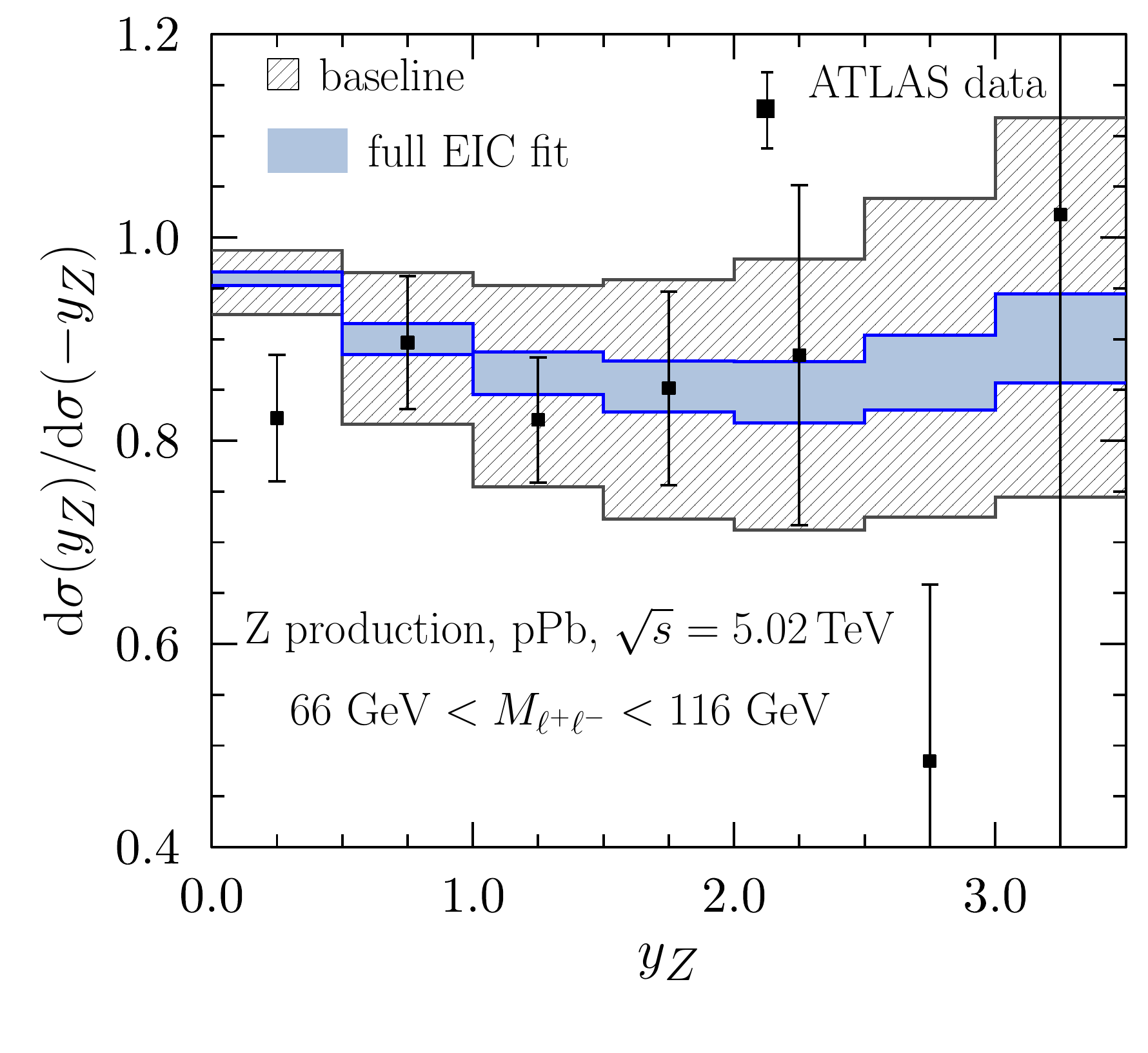}
\includegraphics[width=0.32\linewidth]{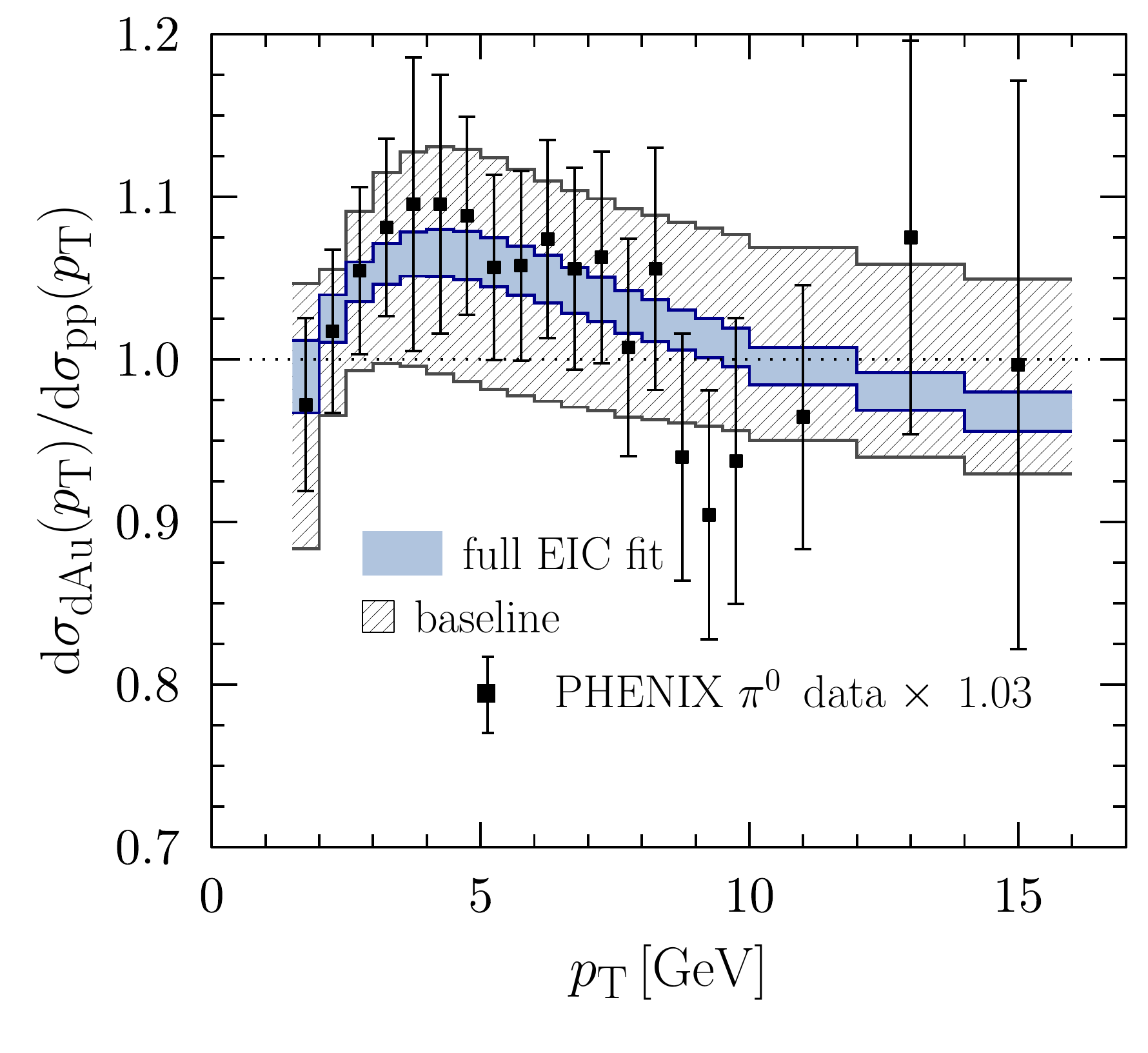}
\includegraphics[width=0.32\linewidth]{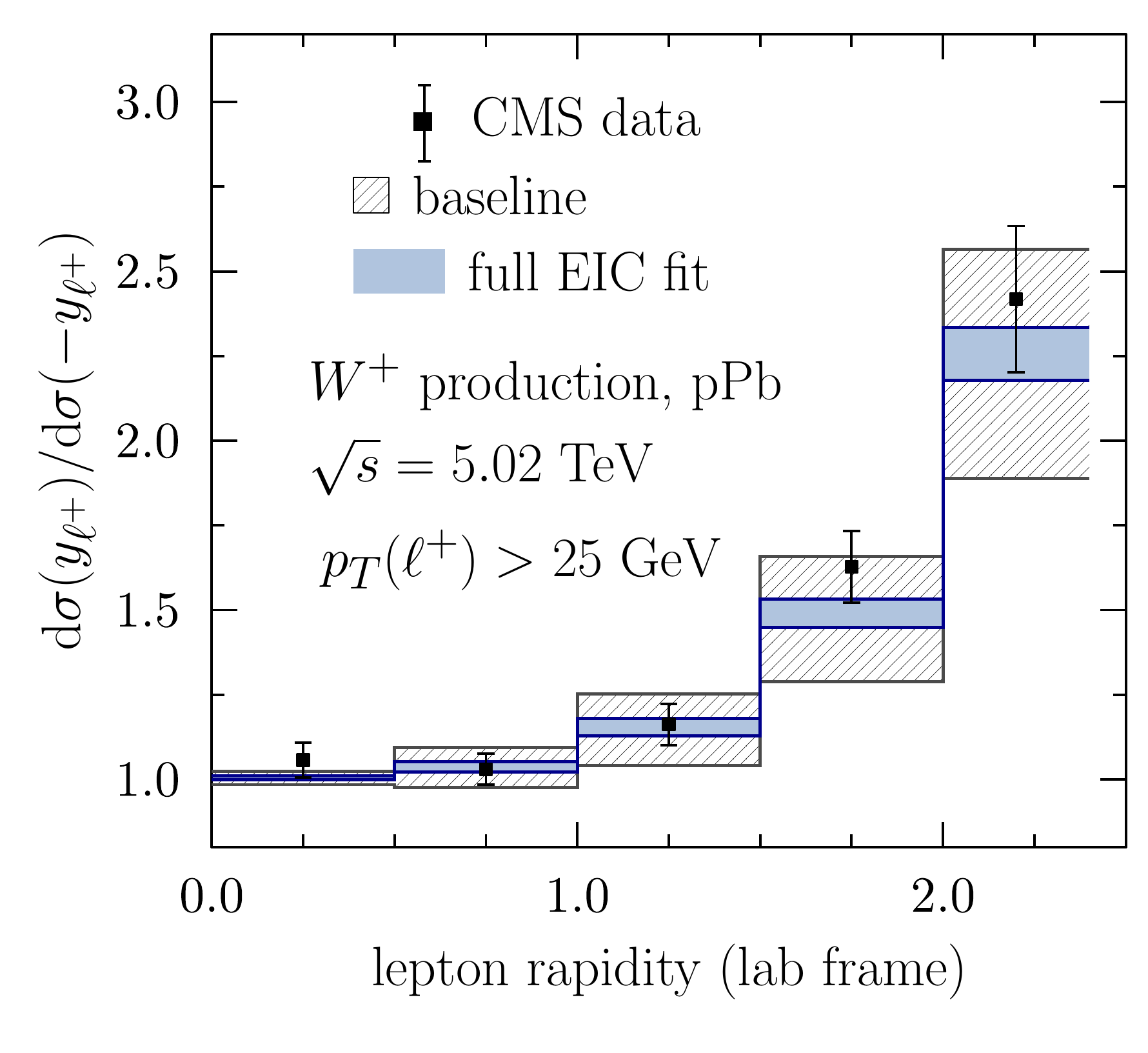}
\includegraphics[width=0.32\linewidth]{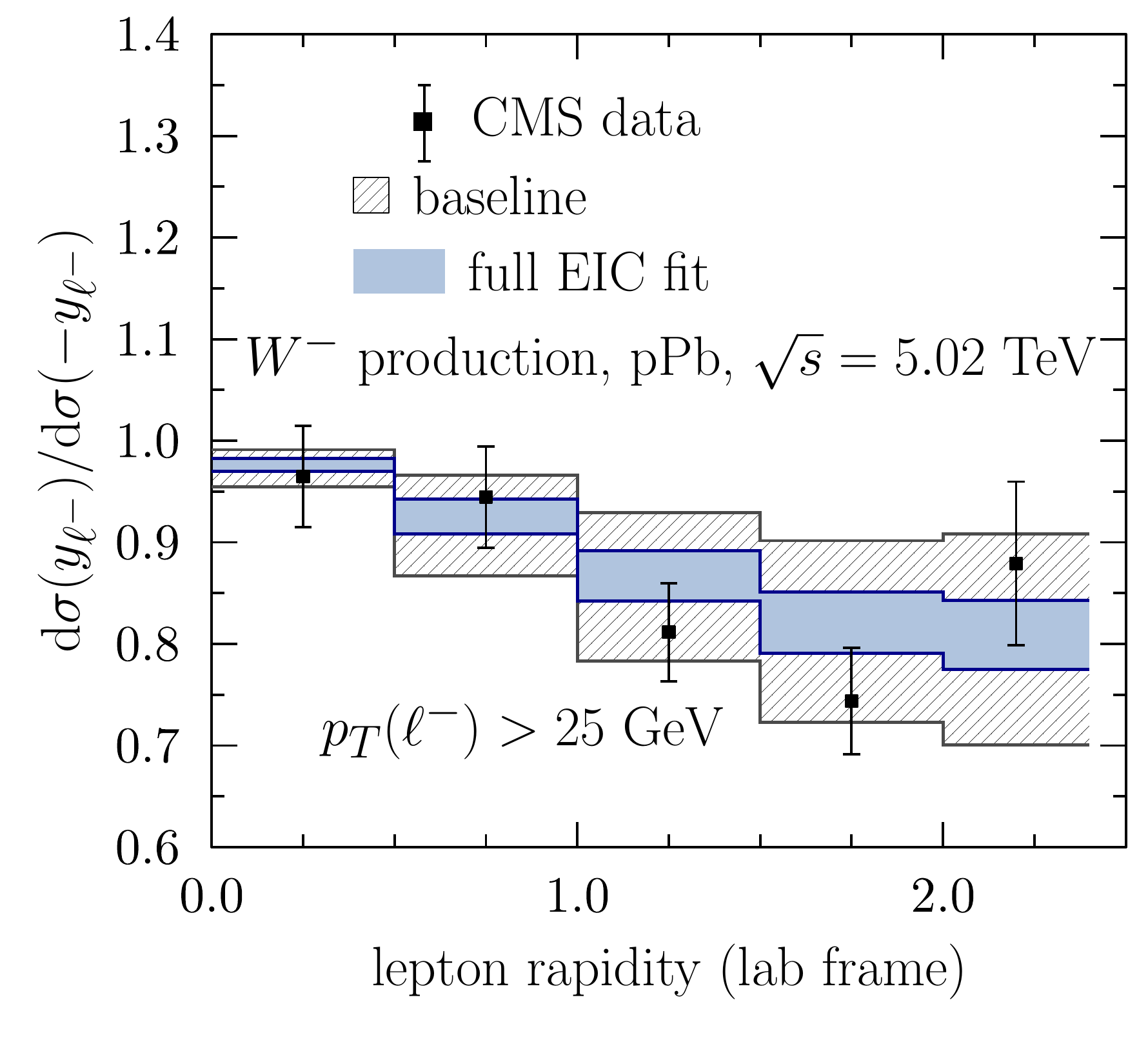}
\includegraphics[width=0.32\linewidth]{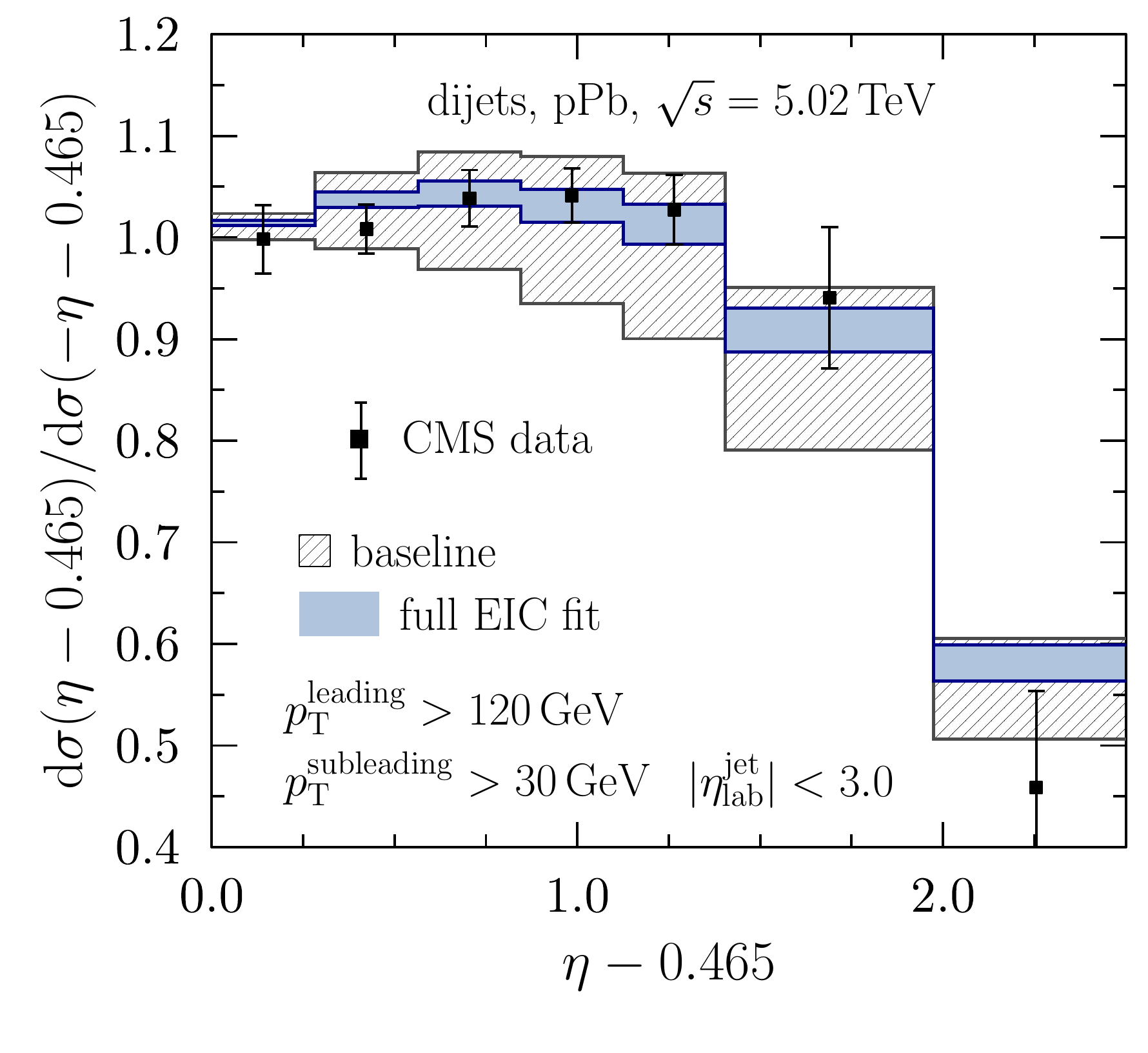}
\caption{The LHC p--Pb and RHIC D--Au data for Z (two upper left panels), ${\rm W}^\pm$ (two lower left panels), dijet production (lower right panel), and inclusive pion production \cite{Adler:2006wg} (upper right panel) compared with the baseline fit (hatched bands) and the full high-energy EIC analysis (blue bands).}
\label{fig:pion}
\end{figure*} 

For the presence of significant anticorrelation, it can be expected that the flavor-averaged nuclear modifications 
for valence quarks
\begin{equation}
 R_V^{\rm Pb}(x,Q^2) \equiv \frac{f^{{\rm proton}/A}_{u_{\rm valence}} + f^{{\rm proton}/A}_{d_{\rm valence}}}{f^{{\rm proton}}_{u_{\rm valence}} + f^{{\rm proton}}_{d_{\rm valence}}},
\end{equation}
and for light sea quarks,
\begin{equation}
 R_S^{\rm Pb}(x,Q^2) \equiv \frac{f^{{\rm proton}/A}_{\overline{u}} + f^{{\rm proton}/A}_{\overline{d}}+f^{{\rm proton}/A}_{\overline{s}}}{f^{{\rm proton}}_{\overline{u}} + f^{{\rm proton}}_{\overline{d}}+f^{{\rm proton}}_{\overline{s}}},
\end{equation}
will be much better constrained. 
We note that these two flavour-independent functions (plus the gluon modification) are what the nPDF fits 
(e.g. EPS09 \cite{Eskola:2009uj}, DSSZ \cite{deFlorian:2011fp}) have traditionally parametrized. In the latest global analyses, nCTEQ15 and EPPS16, this practice has been abandoned as being too restrictive. Presumably this will be the case for all the future global fits of nPDFs. 
The results are shown in 
Figure \ref{fig:fit2}, which presents the flavor-averaged quark nuclear modifications. For $R_V$, the differences between 
the baseline and EIC fits remain always quite modest. We recall that the rather small uncertainty at small $x$ is a pure parametrization bias as the functional form was made more flexible only for gluons. For $R_S$ the impact of EIC data is larger, especially at small $x$, $Q^2$ above the parametrization scale. In fact, for the baseline fit, the 
uncertainty is clearly larger at $Q^2=10\,{\rm GeV}^2$ than at the parametrization scale. This results from the 
very large gluon uncertainty in the baseline fit at $Q^2=1.69\,{\rm GeV}^2$, which 
partially transmits to sea quarks through the partonic scale evolution.
Therefore, the effect of EIC pseudodata is to suppress the small-$x$ uncertainty of $R_S$ up to a factor of four at $Q^2=10\,{\rm GeV}^2$, even though the improvement is less sizable at the parametrization scale $Q^2=1.69\,{\rm GeV}^2$.
However, it should be kept in mind that, similarly to the case of $R_V$, the small-$x$ uncertainties of $R_S$, 
particularly at the parametrization scale $Q^2=1.69\,{\rm GeV}^2$, are artificially small due to the stiff original 
functional forms.

\subsection{Impact on theoretical predictions}
Henceforth the impact of the significantly improved nPDF uncertanties on theoretical predictions of experimental observables is discussed.
Figure~\ref{fig:datacomp1} shows examples of the ratios of inclusive reduced cross-sections $\sigma_{\rm r}{(e^-+{\rm C})}/\sigma_{\rm r}{(e^-+{\rm p})}$ and $\sigma_{\rm r}{(e^-+{\rm Au})}/\sigma_{\rm r}{(e^-+{\rm p})}$ for $\sqrt{s} \approx 89.4 \, {\rm GeV}$. They are compared with the predictions from the baseline fit, and with both the fits using only the low-energy pseudodata, and including also the high-energy EIC pseudodata. In comparison to the baseline fit, the low-energy EIC fit leads to clearly reduced uncertainties at small $x$. The inclusion of high-energy data reduces the uncertainties further by another factor of two at smallest values of $x$. Towards large values of $x$, the impact of EIC pseudodata gradually decreases as the constraints from the old fixed-target data start to dominate.

As already discussed, the inclusion of charm-tagged cross sections clearly improves the gluon constraints at large $x$. While the charm data nominally reaches equally small values of $x$ as the inclusive data, the produced charm quarks always originate from ``parent'' gluons (via $g\rightarrow c\overline{c}$ splittings) with clearly higher $x$. Furthermore, the charm measurements range up to $x \sim 0.3$.
Thus, it is not surprising that it is predominantly the large-$x$ region for gluons that gets better determined by the charm data. Examples of the ratios of charm reduced cross sections corresponding to $\sqrt{s} \approx 89.4 \, {\rm GeV}$ are shown in Figure~\ref{fig:datacomp2}. 
The data are compared with the baseline fit including only inclusive data, and the full analysis with the charm data. 
The baseline-fit errors (hatched bands) clearly exceed the estimated data uncertainties and already the addition of inclusive EIC data reduces the uncertainties quite a bit (gray bands). The inclusion of charm data shrinks the uncertainties further especially at large $x$ (blue bands). 

While the jet production at the LHC is known to constrain high-$x$ gluons \cite{Paukkunen:2014pha}, it is unlikely that a precision like the one obtained here could be reached. Potential constraints on nPDFs through jet production at an EIC have been recently investigated \cite{Klasen:2017kwb}. However, at large-$x$, the jets in DIS originate predominantly from valence quarks.
This is in contrast to the charm cross sections in which the contributions of valence quarks start to appear only at
next-to-next-to leading order in pQCD.
Thus, the charm production will be one of the key measurement for the large-$x$ gluons and will shed light on the size of intrinsic charm component in heavy nuclei.

The EPPS16 analysis is currently the only available parametrization to include constraints from the LHC Run-1 p+Pb data. We would like to point out that there is a 
significant complementarity between these LHC measurements and measurements at an EIC. To illustrate this point we present in Figure \ref{fig:pion} the LHC p+Pb data on W \cite{Khachatryan:2015hha}, Z \cite{Khachatryan:2015pzs,Aad:2015gta}, and dijet \cite{Chatrchyan:2014hqa} production included in the EPPS16 fit (thereby also in the fits presented here). The inclusive pion production data measured by PHENIX \cite{Adler:2006wg} at RHIC are shown as well. The data are compared with the baseline fit and the full high-energy EIC analysis. The reduction of the uncertainties upon including the EIC pseudodata is quite dramatic and concretely demonstrates how an EIC and the LHC can complement each other. This is important in order to truly and precisely address the universality of nPDFs.
It should be stressed that the $Q^2$ in typical 
LHC p+Pb observables is much higher than the ones probed at an EIC. Thus, the constraints on observables at low $Q^2$ from these LHC measurements are only scarce. The theoretical uncertainties on LHC observables that probe the low-$Q^2$ and low-$x$ domain (e.g. open charm, exclusive J/$\Psi$) are always bound to be large and in order to obtain reliable constraints a DIS experiment like an EIC is crucial.

%% file: MasterTexFiles/Conclusions.tex
\section{Summary}
\label{sec:Summary}

We have studied the inclusive and charm cross-section measurements at an EIC and especially their impact on the global 
in the framework of EPPS16 using projected pseudodata for an EIC. A special attention was paid on the determination of gluon densities for which an extended small-$x$ parametrization was used. It was shown that an EIC will have an enormous impact on the global extractions of nPDFs, particularly on the gluon which is currently only weakly constrained.
At low resolution scale $Q^2=1.69 \, {\rm GeV}^2$ the gluon distribution can be determined well down to 
$x\sim 10^{-2}$ but towards higher $Q^2$, the small-$x$ uncertainties quickly shrink across all the small-$x$ domain.
The inclusion of the charm-tagged cross-section measurements decreases the gluon uncertainties substantially at 
large $x$. For the quark sector our study is somewhat more limited as our current analysis methodology does not 
permit to use more flexible parametrizations for all the quark flavors simultaneously.
Despite this limitation, our results 
indicate, especially at $Q^2$ above the parametrization scale, a significant reduction of the sea-quark uncertainties.

The high precision and the wide kinematic coverage in $x$ and $Q^2$
achievable for different observables at an EIC will 
allow for stringent tests of the nPDF universality. Ultimately, such endeavor requires a combination of 
complementary results from the LHC, RHIC and elsewhere.